\documentclass[pdflatex,sn-nature,iicol]{sn-jnl}

\usepackage{graphicx}%
\usepackage{multirow}%
\usepackage{amsmath,amssymb,amsfonts}%
\usepackage{amsthm}%
\usepackage{mathrsfs}%
\usepackage[title]{appendix}%
\usepackage{xcolor}%
\usepackage{textcomp}%
\usepackage{manyfoot}%
\usepackage{booktabs}%
\usepackage{algorithm}%
\usepackage{algorithmicx}%
\usepackage{algpseudocode}%
\usepackage{listings}%
\usepackage{epstopdf}
\usepackage[inkscapelatex=false]{svg}

\usepackage[normalem]{ulem}

\usepackage{breqn}
 
\usepackage{balance}

\definecolor{Lucas}{HTML}{8B008B}      
\definecolor{Arthur}{HTML}{556B2F}   
\definecolor{Raqueti}{HTML}{000080}        

\usepackage{booktabs}   
\usepackage{tabularx}   
\usepackage{siunitx}    
\usepackage{makecell}
\usepackage{array}
\usepackage{amsmath}

\usepackage{cuted}

\usepackage{lmodern}
\usepackage{fix-cm}
\usepackage{mathrsfs}

\begin{document}

\title[Article Title]{Energy Localization in Damped Nonlinear Disordered Metastructures under Superharmonic Resonance}


\author*[1,2]{\fnm{Lucas} \sur{José Dantas Alcântara}}\email{Lucasjosedantas@usp.br}

\author[1]{\fnm{Arthur} \sur{Silva Barbosa}}\email{arthur.silva.barbosa@usp.br}

\author[2]{\fnm{Rafael} \sur{da S. Raqueti}}\email{rafael.dasilva@femto-st.fr}

\author[2]{\fnm{Najib} \sur{Kacem}}\email{najib.kacem@femto-st.fr}
  
\author[1]{\fnm{Leopoldo} \sur{Pisanelli Rodrigues de Oliveira}}\email{leopro@sc.usp.br}

\author[2]{\fnm{Noureddine} \sur{Bouhaddi}}\email{noureddine.bouhaddi@femto-st.fr}

\affil[1]{\orgname{University of São Paulo}, \orgdiv{Department of Mechanical Engineering, São Carlos School of Engineering}, \orgaddress{\street{Av. Trabalhador São-carlense 400}, \city{São Carlos}, \postcode{13566-590}, \state{São Paulo}, \country{Brazil}}}

\affil[2]{\orgname{Université Marie et Louis Pasteur, CNRS, institut FEMTO-ST}, \orgaddress{\city{F-25000 Besançon}, \country{France}}}


\abstract{
This paper proposes a framework for energy localization in nonlinear oscillator chains operating in non-fundamental resonance, with emphasis on superharmonic regimes. 
The system is modeled as a metastructure composed of Duffing oscillators with both linear and nonlinear coupling, incorporating disorder-induced periodicity breaking. Under the assumption of strong excitation, the method of multiple scales is employed to derive the governing equations for soliton dynamics. At first-order perturbation, the classical form of the Nonlinear Schrödinger Equation emerges, whereas second-order analysis yields a previously unreported equation arising from the restitution of time scales.
Analytical and numerical results demonstrate the nucleation of solitons in both hardening and softening regimes, based on two approaches: direct time-domain simulations from initially motionless states and numerical continuation in the frequency domain. A key finding is the distinct role of phase in superharmonic resonances compared to the primary resonance; specifically, the coexistence of multiple frequency components in the steady-state response precludes interpreting the soliton directly as a displacement envelope. Instead, the resulting secular terms captures the soliton associated with the resonant contribution, while transient components remain present under superharmonic excitation. Furthermore, robustness against disorder uncertainty is assessed by determining the tolerance levels that preserve the phenomenon.
These results support the development of vibration control strategies aimed at mitigating the increase in resonant frequencies associated with the geometric downscaling of mechanical systems.
}

\keywords{Nonlinear metastructures, Energy localization, Superharmonic resonance, Standing solitons, Disordered Nonlinear Schrödinger Equation}

\maketitle

\clearpage
\section{Introduction}\label{Intro}

The scientific community has increasingly focused on investigating novel strategies for energy management in mechanical systems \cite{Wu2021}. In fact, the development of metamaterials and metastructures has emerged as a promising approach for addressing a broad range of challenges, including wave propagation \cite{doi:10.1177/1045389X221147668}, vibration control \cite{DASRAQUETI2025119209}, impact absorption/mitigation \cite{Das2024}, sensing \cite{Hu2024}, acoustic protection \cite{SAMPAIO2021116374}, and energy harvesting \cite{Schimidt2023}. Such mechanisms have garnered prominence due to their versatile applications \cite{lee2022piezoelectric} and tunability \cite{sun2020tunable}, allowing dual functionalities and multiphysics couplings \cite{xu2025integrating}. Notably, metastructured lattices enable enhanced energy control in multiphysical systems through collective dynamic effects. For instance, energy localization has been demonstrated in coupled-beam systems \citep{liu2019amplitude}, phononic crystals and topological beams \citep{lu2022vibration}, as well as in defect-induced modes \citep{cao2022defect}, acoustic black holes \citep{yan2023study}, and architected geometries with instability-driven localization \citep{cao2024buckling}.

With respect to the design of metastructures for vibration-related applications, the literature reports a broad spectrum of methodological approaches, encompassing post-buckling instability \cite{jiao2020mechanical}, energy absorption \cite{liu2024never}, topology optimization \cite{lu2025topology}, voxel-based modeling \cite{abhishek2024reversible}, bio-inspired architectures \cite{li2024novel}, wave-based finite element formulations \cite{MDESSANTOS2025104314}, stochastic modeling frameworks \cite{santos2025stochastic}, and inverse design strategies \cite{li2025cvae}, among others. By introducing nonlinear effects into the dynamics of such systems, a new variety of physical phenomena can be explored \cite{Patil2021}. In a recent contribution, Bai et al. (2025) \cite{Bai2025} reviewed the major advances in nonlinear vibration metamaterials. Building upon the classification proposed therein, these advances may be broadly grouped according to the prevailing nonlinear mechanisms, including granular media–induced nonlinearity \cite{espinosa2024pulse}, nonlinear stiffness–based architectures (encompassing quadratic, cubic, negative, and quasi-zero stiffness) \cite{lin2025enhanced}, multistable and bistable systems exploiting snap-through and hysteresis \cite{CHEN2023111053}, as well as designs that explicitly exploit nonlinear dissipation \cite{WANG2019167}. Such strategies have resulted in, among other effects, amplitude-dependent bandgaps \cite{Ganesh2013821}, broadband low-frequency attenuation \cite{SHENG2021115739}, and nonreciprocal wave propagation \cite{gong2025band}.

Within the same scenario, coupled nonlinear oscillators have been widely investigated owing to their rich dynamical behavior \cite{polczynski2021nonlinear, sone2022topological, lenci2022exact, bukhari2023breather,azizi2025bifurcation}. Unlike linear chains, the nonlinear ones can support Intrinsic Localized Modes (ILMs) in perfectly periodic configurations, without the need for disorder or defects \cite{sato2008visualizing, sato2011experimental,BARBOSA2024111358,BARBOSA2025134612}, although the energy localization induction is significantly enhanced by deliberate periodicity breaking \cite{barbosa2024standing}. When addressing studies related to ILMs, a complementary research direction emerges, providing the mathematical framework of energy localization in nonlinear oscillator chains, namely, the investigation of stationary solitons. Following the early observation of solitary waves by Russell \cite{russell1845report} and their subsequent theoretical formalization \cite{korteweg1895xli}, solitons have become fundamental features across a broad range of physical systems. Their dynamics have been extensively described from the standpoint of the Nonlinear Schrödinger Equation (NLS), which has been employed to model phenomena in diverse fields \cite{boopathy2022nonlinear,PhysRevApplied.20.014012,shehzad2023multi,hernandez2022soliton}, and, more recently, in mechanical and elastic metamaterials \cite{ali2022filamentation}. From the perspective of mechanical metastructure architecture, ILMs/stationary solitons have been reported in a broad class of systems, including micromechanical arrays \cite{kenig2009intrinsic}, coupled pendula \cite{jallouli2017}, rotors \cite{fontanela2019dissipative}, and linear oscillator networks \cite{adile2021dynamics}. Their properties and practical applications have been investigated in the literature, particularly in sensing \cite{grenat2022mass}, sensitivity enhancement \cite{manav2018mode}, and nonlinear energy sink technologies \cite{vakakis2014interactions}.

Recently, Gao et al. (2024) \cite{gao2024brief} published a comprehensive review classifying the state of the art in solitary waves within nonlinear metamaterials and metastructures. The reported examples encompass pulse-type solitons \cite{wang2020influencing}, kink-type solitons \cite{librandi2021programming}, envelope-type solitons \cite{barbosa2022methodological}, breathers \cite{li2017experimental}, vector solitons \cite{deng2019focusing}, rarefaction solitons \cite{deng2019propagation}, and topological solitons \cite{yasuda2020transition}, among others. The reviewed contributions highlight the dynamical advantages of nonlinear wave phenomena over linear ones, particularly regarding robustness against waveform dispersion \cite{yu2025propagation}. Consistent with such advantage, the pursuit of self-balanced states (in which energy remains predominantly confined to a region of the space) can be induced by non-propagating solitary waves, an idea that has been established in the literature for several years  \cite{mokhtari2013numerical}. Studies have derived closed-form analytical solutions and the corresponding stability maps for standing/stationary solitons under parametric  \cite{kenig2009pattern, barashenkov1991stability} and external \cite{BARBOSA2024111358,barashenkov1996existence} excitation. Extending beyond periodic configurations, quasiperiodic lattices with localized impurities have been shown to promote multi-soliton nucleation and to expand the existence domain of damped solutions \cite{alexeeva2000impurity, barbosa2024standing}. Furthermore, the nature of the excitation plays a critical role in solitonic dynamics, and analytical solutions for stationary damped solitons subjected to external forcing remain unavailable \cite{BARBOSA2023110879, barashenkov1996existence}. In contrast to other physical domains, the aforementioned studies add a further layer of complexity to the problem, as damping is generally non-negligible in mechanical systems. In \cite{partI,partII}, Barbosa et al. (2026) proposed a design methodology grounded in the Disordered Damped Nonlinear Schrödinger Equation, demonstrating the feasibility of using standing solitons as design guidelines for energy harvesting applications. The authors further highlight the challenges posed by damping, particularly with respect to the generation of solitons in the frequency domain.

Despite the extensive literature on the topic, to the best of the authors’ knowledge, several nonlinear phenomena involving non-fundamental harmonic resonances and metastructures have attracted limited research effort. Indeed, resonances not associated with the fundamental frequency have been documented for decades \cite{Abe1998,Fahsi2009} and continue to attract attention in recent studies \cite{Raze2025,Raze2025a,maharshi2025nonlinear,yuan2025vibration}. Applications such as pedestrian-induced excitation of vibrating structures \cite{Hu2025}, energy harvesting exploiting superharmonic behavior \cite{Rezaei2020}, MEMS dynamics \cite{Sarafraz2019}, multi-staged clutch damper systems \cite{Yoon2022}, time-delayed feedback control \cite{Peng2019}, and damage regulation \cite{zhou2024locally} further motivate the investigation of secondary resonances in nonlinear oscillators. From the perspective of metastructure development under secondary resonance regimes, the study of coupling among unit cells becomes of particular interest \cite{Jothimurugan2015,lepidi2019wave}, raising questions related to energy management under such conditions. For instance, in \cite{wu2023nonlinear}, Wu et al. investigated the generation of superharmonic waves in a chain-type multicell aperiodic structure with nonlinear stiffness. In a later development, Lepidi et al. \cite{Lepidi2025} presented a study on superharmonic wave propagation in two-dimensional mechanical metamaterials with inertia amplification. Beyond superharmonic responses, subharmonic resonances have been reported in locally resonant metamaterials induced by autoparametric resonance \cite{silva2019emergent}, in two-dimensional wings incorporating bistable metamaterials \cite{hu2024effectively}, and in piezoelectric metamaterial beams subjected to viscous flow \cite{liu2025nonlinear}. Nevertheless, as far as the authors are aware, there are no studies addressing the energy localization in nonlinear metastructures operating outside the primary resonance regime from solitons stand point.

Given the current state of the art, this paper aims to demonstrate the existence of stationary solitons in secondary resonance regimes, particularly superharmonic ones. To this end, a theoretical mechanical metastructure is investigated, and, based on the method of multiple scales under the strong excitation assumption \cite{shen2020primary,chen2025primary}, a discrete soliton equation is derived. Motivated by a recent study \cite{partI,partII}, in which the behavior of multiphysics solitons was assessed in the frequency domain, the present work establishes parallels between the mechanisms associated with primary and superharmonic resonances, highlighting their implications for energy localization. The main contributions of this article consist of (i) demonstrating the role of phase dynamics in lattice energy localization, (ii) describing the wave nucleation process from initially motionless oscillators and (iii) derive a novel control methodology for energy localization in nonlinear lattices under disorder uncertainty; contributions that have so far been addressed in the literature only around the fundamental frequency. Given the generalizing character of the theoretical formulation, which is not restricted to a specific metastructure architecture or to a particular type of nonlinearity, the analyses are conducted for both hardening and softening behavior, thereby clarifying the numerical differences between the two scenarios from the perspective of the frequency domain\footnote{For didactic consistency, the same color schemes are adopted throughout the study for the hardening (``viridis'') and softening (``plasma'') regimes.}. We expect this work to contribute to the field of metastructures by proposing a new design paradigm that leverages the robustness of solitons against waveform dispersion. By establishing the viability of solitary waves operating under superharmonic resonance, the realization of systems with reduced geometric dimensions becomes a more attainable prospect, as the increase in natural frequency associated with geometric miniaturization can be offset by the availability of harmonic frequencies, which may act as submultiples of the fundamental one and accordingly broaden the design space. 

The remainder of the paper is structured as follows. Section \ref{Derivation of the lattice soliton equation in the superharmonic regime} presents the derivation of the lattice soliton equation in the superharmonic regime, based on the method of multiple scales. The subsection on first-order secular terms establishes the connection with the Disordered Damped Nonlinear Schrödinger equation, while the subsection on second-order secular terms introduces a novel restitution equation that accounts for higher-order effects. Section \ref{Duffing oscillator frequency response in superharmonic regimes} investigates the frequency response of an isolated Duffing oscillator under superharmonic excitation, providing a reference for interpreting the collective behavior of the lattice. Section \ref{Energy nucleation and stationary soliton analysis} analyzes the nucleation and properties of stationary solitons, including the influence of disorder, the saturation with respect to the number of oscillators, and the uncertainty levels tolerated by the metastructure. Finally, Section \ref{Concluding remarks} summarizes the main findings and discusses perspectives for future work.

\section{Derivation of the lattice soliton equation in the superharmonic regime}\label{Derivation of the lattice soliton equation in the superharmonic regime}

\begin{figure}[t]
    \centering
    \includegraphics[width=0.36\textwidth]{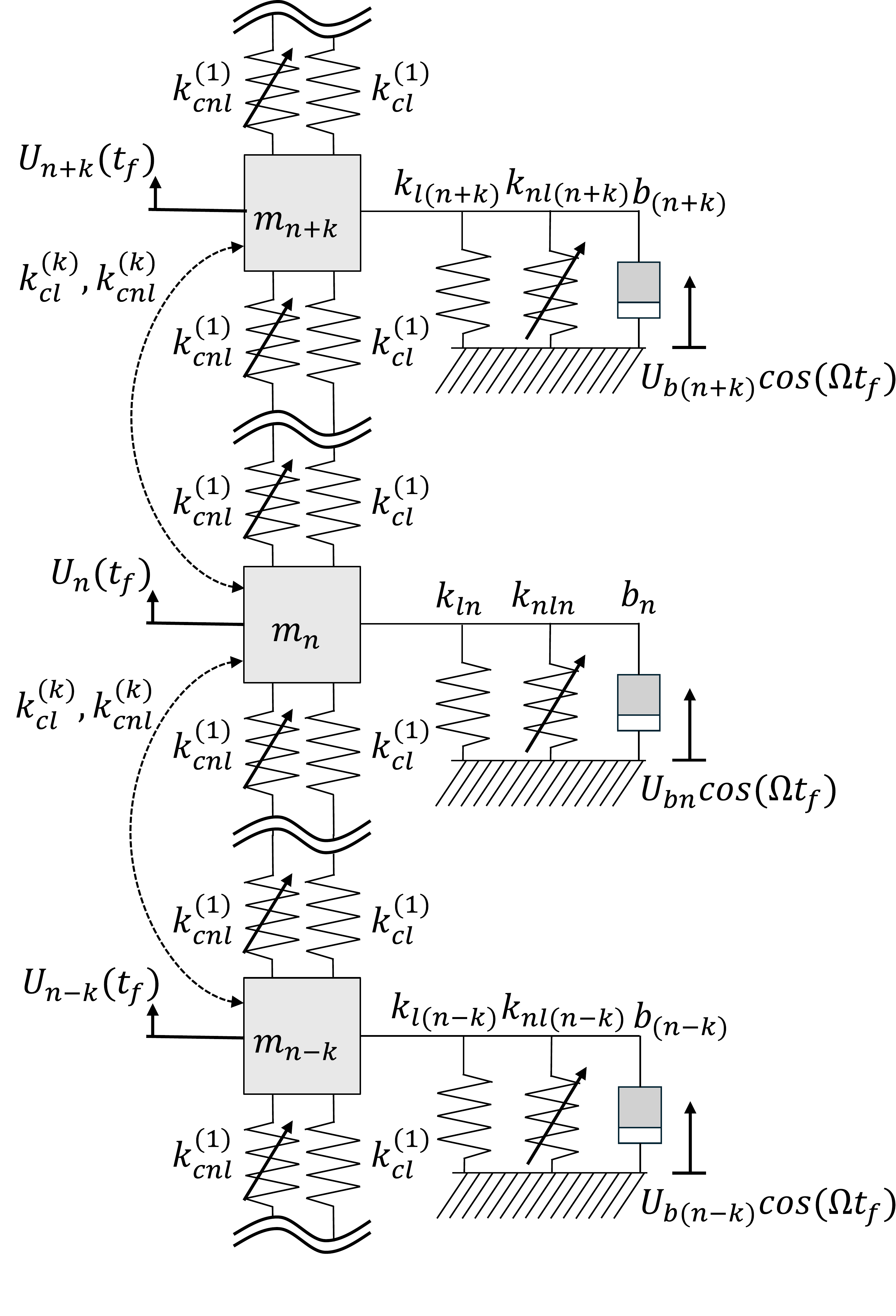}
    \caption{Equivalent lumped-parameter model of the metastructure. The system consists of a chain of oscillators coupled through linear stiffnesses $k_{cl}$ and nonlinear stiffnesses $k_{cnl}$, with coupling interactions extending up to the $\alpha$-th neighbor and the superscript $^{(k)}$ denotes the coupling order. Each unit cell is defined by the mass $m$, linear stiffness $k_l$, nonlinear stiffness $k_{nl}$, damping coefficient $b$, and base excitation of amplitude $U_b$, indexed by $n$ to identify each oscillator. The excitation is tuned around third order superharmonic resonance of the primary mode.}
    \label{Sistema}
\end{figure}

The goal of this section is to present the mathematical description of the steady-state regime of the nonlinear metastructure shown in Fig. \ref{Sistema}. Based on the resulting equation, the differences relative to primary resonance regimes are detailed, highlighting theoretical features that arise exclusively in secondary resonances, such as the emergence of multiple frequency components in the composition of the response. Additionally, the NLS is derived at the first slow temporal scale, thus demonstrating the possibility of superharmonic soliton generation. To extend the validity of the model, a higher temporal order is incorporated in the derivation of the secular terms, resulting in an equation not yet reported in the literature.

By enforcing force equilibrium among the $N$ components illustrated in Fig. \ref{Sistema}, the governing equation of the lattice dynamics under base acceleration can be written as follows:
\begin{multline}
\dfrac{d^{2} U_n(t_f)}{d t_f^{2}}
+ 2  {\zeta_n} \omega_0 \dfrac{d U_n(t_f)}{d t_f}
+ \omega_0^{2} U_n(t_f)\\
+ f(n) U_n(t_f) 
\pm \beta  U_n^{3}(t_f)\\
+ \sum_{k=1}^{\alpha} C_l^{(k)} \Big( 2 U_n(t_f) - U_{n+k}(t_f) - U_{n-k}(t_f) \Big)\\
+ \sum_{k=1}^{\alpha} C_{nl}^{(k)} \Big[ \big(U_n(t_f) - U_{n+k}(t_f)\big)^{3} \\+ \big(U_n(t_f) - U_{n-k}(t_f)\big)^{3} \Big]
= -  {F_n} \cos (\Omega t_f),
\label{Equação mecânica dim}
\end{multline}where the natural frequency $\omega_0$, the damping ratio $\zeta_n$, the spatial variation of the linear stiffness $f(n)$, the nonlinear stiffness $\beta$, the linear couplings $C_l^{(k)}$, and the nonlinear couplings $C_{nl}^{(k)}$ are defined in Table \ref{parâmetros_lattice}. It is important to emphasize that the system incorporates couplings not only among neighboring oscillators but also among oscillators separated by up to order $\alpha$. Additionally, periodicity breaking is allowed in the damping ratio, linear stiffness, and oscillator excitation, while the couplings and nonlinear effects remain periodic.

In order to address Eq. \eqref{Equação mecânica dim}, the method of multiple scales is applied \cite{nayfeh2008nonlinear}. The lattice physical parameters are therefore rescaled by a factor $\varepsilon$ according to the following orders\footnote{For notational convenience, the newly introduced variables are defined using the same notation as in Eq. \eqref{Equação mecânica dim}, indexed by $\varepsilon$.}:\begin{multline}
\dfrac{d^{2} U_n(t_f)}{d t_f^{2}}
+ \omega_{0}^{2} U_n(t_f)\\
+ \varepsilon
\Bigg[
2 {\zeta_n}_{\varepsilon} \omega_{0} \dfrac{d U_n(t_f)}{d t_f}
+ U_n(t_f) f_{\varepsilon}(n)
\pm \beta_{\varepsilon} U_n^{3}(t_f)\\
+ \sum_{k=1}^{\alpha} C_{l\varepsilon}^{(k)}
\left( 2 U_n(t_f) - U_{n+k}(t_f) - U_{n-k}(t_f) \right)
\Bigg]\\
+ \varepsilon^{2}
\sum_{k=1}^{\alpha} C_{nl\varepsilon^2}^{(k)}
\Bigg[
\left( U_n(t_f) - U_{n+k}(t_f) \right)^{3}\\
+ \left( U_n(t_f) - U_{n-k}(t_f) \right)^{3}
\Bigg]
= - {F_n} \cos(t_f \Omega).
\label{Equação mecânica dim pertubada}
\end{multline} As indicated by Eq. \eqref{Equação mecânica dim pertubada}, the high-excitation hypothesis is adopted, under which the variable $F_n$ is not perturbed \cite{10.1115/1.4006183}. Additionally, the linear and nonlinear coupling stiffnesses are assumed to be one order of magnitude lower than the corresponding linear and nonlinear stiffnesses, respectively. Given the extensive algebraic manipulations associated with the method of multiple scales, Eq. \eqref{Equação mecânica dim pertubada} is nondimensionalized according to the relations presented in Table \ref{parâmetros_lattice}, from which the following governing equation is derived:\begin{multline}
\dfrac{d^{2} u_n(t)}{d t^{2}}
+ u_n(t)
+ \varepsilon
\Bigg[ 2  {\zeta_n}_{\varepsilon} \dfrac{d u_n(t)}{d t} \pm u_n^{3}(t)\\
+ \sum_{k=1}^{\alpha} \chi_{l}^{(k)} \Big( 2 u_n(t) - u_{n+k}(t) - u_{n-k}(t) \Big)\\
+ u_n(t) \mu(n)
\Bigg]
+ \varepsilon^{2}
\sum_{k=1}^{\alpha} \chi_{nl}^{(k)}
\Bigg[
\Big( u_n(t) - u_{n+k}(t) \Big)^{3}\\
+ \Big( u_n(t) - u_{n-k}(t) \Big)^{3}
\Bigg]
= -{K_n} \cos( t \Theta ).
\label{Eq mec normalizada}
\end{multline} In Eq. \eqref{Equação mecânica dim}, the nonlinear coefficient $\beta$ is defined as positive, while hardening and softening behaviors are represented by the sign $\pm$. Such convention is used consistently in the algebraic development.

Hassan \cite{hassan1995second} demonstrated that, in the third superharmonic regime of the Duffing oscillator, at least a second-order perturbation expansion is required to properly capture nonlinear harmonic interactions and yield quantitatively reliable steady-state predictions. This requirement arises because, when only a first-order perturbation is considered, the fundamental harmonic acts as an ideal energy source for the third-order superharmonic, such that the amplitude of the superharmonic does not influence the fundamental component of the response, leading to inaccurate predictions when this assumption is no longer valid. In light of this result, the solution of Eq. \eqref{Eq mec normalizada} is expanded up to the second order as follows:\begin{multline}
u_n(t) = u_n^{0}(T_0,T_1,T_2)
+ \varepsilon\,u_n^{1}(T_0,T_1,T_2) \\
+ \varepsilon^2\,u_n^{2}(T_0,T_1,T_2)
+ \mathcal{O}(\varepsilon^3),
\label{solução}
\end{multline}where $T_j = \varepsilon^{j} t$, for $j = 0,1,2$, with secular terms up to $\varepsilon^2$. Owing to the mathematical ansatz in Eq. \eqref{solução}, a hierarchy of linear equations in $\varepsilon$ powers is obtained, as follows\footnote{For notational compactness, $u_n^{0}(T_0,T_1,T_2) = u_n^{0}$ and $u_n^{1}(T_0,T_1,T_2) = u_n^{1}$.}:

\begin{equation}
\mathcal{O}(\varepsilon^0): \ \dfrac{\partial^{2} u_n^{0}}{\partial T_{0}^{2}}
+ u_n^{0}
= - {K_n} \cos\left(T_{0} \Theta \right),
\label{potencia_0}
\end{equation}

\begin{multline}
\mathcal{O}(\varepsilon^1): \ 
 \dfrac{\partial^{2} u_n^{1}}{\partial T_{0}^{2}} + u_n^{1} = \\
-\Bigg(
\pm \left(u_n^{0}\right)^{3}
+ 2  {\zeta_n}_{\varepsilon} \dfrac{\partial u_n^{0}}{\partial T_{0}}
+ 2 \dfrac{\partial^{2} u_n^{0}}{\partial T_{1} \partial T_{0}}
+ u_n^{0} \mu(n)\\
+ \sum_{k=1}^{\alpha} \chi_l^{(k)}
\Big( 2 u_n^{0} - u_{n+k}^{0} - u_{n-k}^{0} \Big)
\Bigg),
\label{potencia_1}
\end{multline} and

\begin{multline}
\mathcal{O}(\varepsilon^2): \ 
\dfrac{\partial^{2} u_n^{2}}{\partial T_{0}^{2}} + u_n^{2} = \\
-\Bigg(
\pm 3 \left(u_n^{0}\right)^{2} u_n^{1}
+ 2  {\zeta_n}_{\varepsilon} \dfrac{\partial u_n^{0}}{\partial T_{1}}
+ 2  {\zeta_n}_{\varepsilon} \dfrac{\partial u_n^{1}}{\partial T_{0}}\\
+ u_n^{1} \mu(n)
+ \dfrac{\partial^{2} u_n^{0}}{\partial T_{1}^{2}}
+ 2 \dfrac{\partial^{2} u_n^{0}}{\partial T_{2} \partial T_{0}}
+ 2 \dfrac{\partial^{2} u_n^{1}}{\partial T_{1} \partial T_{0}} \\
+ \sum_{k=1}^{\alpha} \chi_l^{(k)} \Big( 2 u_n^{1} - u_{n+k}^{1} - u_{n-k}^{1} \Big)\\
+ \sum_{k=1}^{\alpha} \chi_{nl}^{(k)}
\Big[
\left(u_n^{0} - u_{n+k}^{0}\right)^3
+ \left(u_n^{0} - u_{n-k}^{0}\right)^3
\Big]
\Bigg).
\label{potencia_2}
\end{multline}

A distinguishing feature from the primary resonance case is that, unlike the situation in which the variable $K_n$ is perturbed, Eq. \eqref{potencia_0} admits a solution composed of two terms, formulated as:\begin{multline}
u_n^{0}\!\left(T_{0},T_{1},T_{2}\right)
= \\ {\Lambda_{n}} e^{i T_{0}\Theta}
+ A_n\!\left(T_{1},T_{2}\right)e^{i T_{0}}
+ \text{c.c.},
\label{solu_u0}
\end{multline}where c.c. denotes the complex conjugate and ${\Lambda_{n}}$ is defined as \begin{equation}
{\Lambda_{n}} = \dfrac{{K_n}}{2\left(\Theta^{2}-1\right)}.
\label{Lambda}
\end{equation}Given the focus on superharmonic resonance, the nondimensional excitation frequency $\Theta$ is defined in terms of a tuning parameter $\sigma$, according to:\begin{equation}
\Theta = \dfrac{1}{3}\left(1 + \varepsilon\sigma\right).
\label{Theta}
\end{equation}Equation \eqref{Theta} constitutes the starting point for the subsequent steps. It specifies the harmonic of interest, from which the analysis can be extended to higher-order subharmonic or superharmonic cases. Although the present study focuses on the formulation given in Eq. \eqref{Theta}, the mathematical treatment can be readily generalized to other particular configurations.

\subsection{First-Order Secular Terms}
\label{First-Order Secular Terms}

Upon substitution of Equations \eqref{solu_u0}, \eqref{Lambda}, and \eqref{Theta} into the first-order Eq. \eqref{potencia_1}, the first-order resonant terms, proportional to $e^{i T_{0}}$, arise and must be cancelled. Accordingly, the following Equation must be respected \footnote{For notational compactness, the time dependence is omitted.}:\begin{multline}
\pm\,{\Lambda^{3}_{n}} e^{i T_{1}\sigma}
 \pm\,3A_n\left|A_n\right|^2 \\
+ \left(
\pm\,6{\Lambda^{2}_{n}}
+ 2 i  {\zeta_n}_{\varepsilon}
+ \mu(n)
\right) A_n
+ 2 i \dfrac{\partial A_n}{\partial T_{1}} \\
+ \sum_{l=1}^{\alpha} \chi_l^{(l)}
\left(
2A_n - A_{n+l} - A_{n-l}
\right) = 0.
\label{Seculares_1}
\end{multline}Equation \eqref{Seculares_1} corresponds to the Discrete NLS, widely reported in the literature \cite{HENNIG1999333,PhysRevA.76.042108,Dai_2008,PhysRevLett.106.078102} and still the subject of ongoing research \cite{ASGHARI20247}. The resulting mathematical framework already enables the use of superharmonic ILMs as mechanisms for controlling energy localization in the structure shown in Fig. \ref{Sistema}. However, unlike solitons derived under primary resonance conditions, higher-order powers of $\varepsilon$ should be considered for a more accurate amplitude quantification \cite{hassan1995second}.

Further developing the algebraic derivation and disregarding the continuity hypothesis, it is verified that, once the first-order secular terms are removed, Equation \eqref{potencia_1} yields the following solution:

{\small
\begin{multline}
u_n^{1}(T_0,T_1,T_2)
=\\
 \pm\dfrac{3{\Lambda_{n}} 
A_{n}^{2}
e^{\dfrac{i \left(5 T_{0} - T_{1} \sigma\right)}{3}}
}{(\Theta - 1)(\Theta - 3)}
\pm \dfrac{ 3{\Lambda^{2}_{n}} 
A_{n}
e^{\dfrac{i \left(5 T_{0} + 2 T_{1} \sigma\right)}{3}}
}{4 \Theta \left(\Theta + 1\right)}
\\[6pt]
\pm \dfrac{3 {\Lambda^{2}_{n}}
A_{n}
e^{\dfrac{i \left(T_{0} - 2 T_{1} \sigma\right)}{3}}
}{4 \Theta \left(\Theta - 1\right)}
\pm \dfrac{3 {\Lambda_{n}} A_{n}^{2}
e^{\dfrac{i \left(7 T_{0} + T_{1} \sigma\right)}{3}}
}{(\Theta + 1)(\Theta + 3)}
\\[6pt]
\pm \dfrac{ 
A_{n}^{3} e^{3 i T_{0}}
}{8}
+ \Bigg(
\dfrac{2 i \Theta {\Lambda_{n}} {\zeta_n}_{\varepsilon}}{\Theta^{2} - 1}
+
\dfrac{{\Lambda_{n}} \mu(n)}{\Theta^{2} - 1}\\
\pm \dfrac{3 {\Lambda^{3}_{n}}}{\Theta^{2} - 1}
\pm
\dfrac{6 {\Lambda_{n}} \left|A_{n}\right|^{2}}{\Theta^{2} - 1}
\Bigg)
e^{ \dfrac{i \left(T_{0} + T_{1} \sigma\right)}{3}}
+ \mathrm{c.c.},
\label{solu_u01}
\end{multline}
}which, together with Eq. \eqref{solu_u0}, constitute the solution proposed in Eq. \eqref{solução}. A further parallel with primary resonance concerns the amplitudes involved in the construction of the solution. As shown, both $A_n$ and $\Lambda_{n}$ must be evaluated in steady state, although the soliton is associated with the localization of $A_n$.

\subsection{Second-Order Secular Terms and Restitution Equation}
\label{Second-Order Secular Terms e Restitution Equation}

In analogy with the derivation of the $\varepsilon$-order secular terms in Eq.\eqref{Seculares_1}, the substitution of \eqref{solu_u01} into the $\varepsilon^2$-order problem in \eqref{potencia_2} leads to additional resonant terms, expressed as:

{\small
\begin{multline}
\mathcal{L}_{1} A_{n}^{2}e^{-iT_{1}\sigma}
+ \left(\mathcal{L}_2(n) + \mathcal{L}_3\left|A_n\right|^{2}\right)e^{iT_1\sigma}\\
+ \Bigg(
\mathcal{L}_{4}(n)
+ \mathcal{L}_{5}(n)\left|A_{n}\right|^{2}
- \dfrac{15\left|A_{n}\right|^{4}}{8}
\Bigg)A_n \\
+ 2i\,\dfrac{\partial A_n}{\partial T_2} 
+\mathcal{F}_{\chi_l^{(k)}}
+\mathcal{F}_{\chi_l^{(k)}\chi_l^{(j)}}
+\mathcal{F}_{\chi_{nl}^{(k)}}
=0,
\label{Seculares_2}
\end{multline}}where the definitions of the variables $\mathcal{L}_i$ ($i = 1,2,3,4$ \textrm{and} $5$) and the coupling functions $\mathcal{F}_{\chi_l^{(k)}}$, $\mathcal{F}_{\chi_{l}^{(k)}\chi_{l}^{(j)}}$, and $\mathcal{F}_{\chi_{nl}^{(k)}}$ are detailed in Table \ref{auxiliares_sec_2}. 

In contrast to the equation presented in \eqref{Seculares_1}, the resonant terms of Eq. \eqref{Seculares_2} do not correspond to a previously reported formulation in the literature and therefore lack established stability diagrams. Precisely, due to the presence of contributions from distinct temporal scales in $u_n(t)$, the emergence of secular terms across these scales requires the formulation of a restitution equation in the nondimensional time $t$, as expressed in the following equation:
 
\begin{equation}
\dfrac{dA_n}{dt} = \dfrac{\partial A_n}{\partial T_0} +\varepsilon\dfrac{\partial A_n}{\partial T_1} +  \varepsilon^2 \dfrac{\partial A_n}{\partial T_2}.
\label{restituição}
\end{equation}Through the transformation $A_{n}(t) e^{i t \varepsilon \sigma} \rightarrow A_{n}(t)$, which removes the oscillatory character of Eq. \eqref{restituição}, one obtains the steady-state governing Equation as:
{\small
\begin{multline}
\varepsilon^2\mathcal{L}_{1} A_{n}^{2} +\varepsilon^2\mathcal{L}_2 \pm\varepsilon{\Lambda^{3}_{n}} + \varepsilon^2\mathcal{L}_3\left|A_n\right|^{2} \\
+ \left(
\pm\,6\varepsilon{\Lambda^{2}_{n}}
+ 2 i \varepsilon {\zeta_n}_{\varepsilon}
+ \varepsilon\mu(n)
- 2\varepsilon\sigma + \varepsilon^2\mathcal{L}_{4} 
\right) A_n \\
+
\left(\pm3\varepsilon+\varepsilon^2\mathcal{L}_{5}\right)\left|A_{n}\right|^{2}A_{n}
- \dfrac{15\varepsilon^2\left|A_{n}\right|^{4}A_n}{8} \\
+\varepsilon\sum_{k=1}^{\alpha} \chi_l^{(k)}
\left(
2A_n - A_{n+k} - A_{n-k}
\right) \\
+\varepsilon^2\mathcal{F}_{\chi_l^{(k)}}
+\varepsilon^2\mathcal{F}_{\chi_l^{(k)}\chi_l^{(j)}}
+\varepsilon^2\mathcal{F}_{\chi_{nl}^{(k)}} + 2i\,\dfrac{d}{dt}A_n = 0,
\label{soliton}
\end{multline}}where the previously defined coupling functions remain unchanged. Additionally, $A_n$ in Eq.\eqref{Seculares_2} depends on the time scales $T_1$ and $T_2$, while in Eq.\eqref{soliton} it depends only on the reconstituted time $t$.

Not being inherently bound to the system depicted in Fig. \ref{Sistema}, Eq. \eqref{Equação mecânica dim} retains a general, architecture-independent character (e.g., pendulums \cite{jallouli2017}, linearly coupled masses \cite{adile2021dynamics}, or circular structures \cite{fontanela2019dissipative}), which is likewise inherited by Eq. \eqref{soliton}. Despite this fact, approximations related to the analytical treatment introduce a discrepancy between both Equations as the acceleration amplitude increases \cite{hassan1995second}. Hence, to preserve physical consistency, the theoretical soliton predictions associated with Eq. \eqref{soliton} are numerically validated using Eq. \eqref{Eq mec normalizada} (see Fig. \ref{validação_soliton_num} in Appendix \ref{Supplementary numerical simulations}).

From a mathematical standpoint, nontrivial aspects can be identified. The first deals with the type of nonlinearity present in Eq. \eqref{soliton}. Although it originates from a Duffing oscillator with cubic nonlinearity, Eq. \eqref{soliton} also exhibits quadratic and quintic nonlinearities, which emerge from the secular terms described by Eq. \eqref{Seculares_2}. Another observation pertains to the influence of ${\Lambda_{n}}$. Qualitatively, since it depends on the external excitation $K_n$ in Eq. \eqref{Lambda}, it is natural to interpret it as an external forcing term in Eq. \eqref{Seculares_1} \cite{barashenkov1996existence}. The nontrivial aspect lies in the role played by $\Lambda$ in conjunction with the periodicity breaking $\mu(n)$. Such mathematical consequence does not arise under primary excitation, where the periodicity breaking in the natural frequency does not couple to the applied excitation.

A major difference between the energy localization around $\omega_0$ when compared to the superharmonic energy localization concerns the importance of the phase associated with $A_n$. Even when considering the approximation $u_n(t) \approx u_n^{0}(T_0,T_1,T_2)$\cite{nayfeh2008nonlinear}, Eq. \eqref{solu_u0} presents terms whose direct algebraic summation cannot be performed. The amplitudes $\Lambda_n$ and $A_n$ must be considered jointly in the composition of the amplitude of $u_n(t)$, although the soliton Eq. \eqref{Seculares_1} addresses only $A_n$.


\subsection{Continuity Hypothesis}
\label{Continuity Hypothesis}

The formulation presented in Eq. \eqref{soliton} is established within a discrete spatial framework. This consideration provides a more general modeling approach than strategies based on the continuity hypothesis. At the same time, by imposing approximations associated with the oscillator spacing $\Delta s$, theoretical results previously reported in the literature can be recovered. Let us begin by assuming the following approximation along the space $s = n \Delta s$:
{\small
\begin{equation}
A_{n} \rightarrow  A(s,t), \quad \text{and}\quad A_{n\pm1} \rightarrow  A(s\pm\Delta s,t).
\label{continuidade}
\end{equation}
}The adopted approximation allows the use of the following expansion:
{\small
\begin{multline}
A_{n\pm1}
\approx
A 
\pm
\Delta s\,\frac{\partial A}{\partial s}
+
\frac{\Delta s^2}{2}\frac{\partial^2 A}{\partial s^2}
+ \mathcal{O}(\Delta s^4),
\label{continuidade2}
\end{multline}
}which results in the transformation of the operator ${\Delta_1}\{X\}$ (see Table \ref{auxiliares_sec_2}) into an operation that maps a discrete function to its corresponding second-order spatial derivative:
{\small
\begin{equation}
{\Delta_1}\{X\} = -\frac{\Delta s^2}{2}\frac{\partial^2 X}{\partial s^2}+ \mathcal{O}(\Delta s^4),
\label{continuidade3}
\end{equation}
}where $X$ denotes a general function, such that:
{\small
\begin{equation}
\left(
2A_n - A_{n+1} - A_{n-1}
\right) \rightarrow  -\frac{\Delta s^2}{2}\dfrac{\partial^2 A\left(s,T_1\right)}{\partial s^2}.
\end{equation}
}

For the sake of theoretical simplification, the couplings between nonadjacent oscillators ($\alpha = 1$, $\chi_l^{(k)} = \chi_l$), nonlinear couplings ($\chi_{nl}^{(k)} = 0$), and periodicity-breaking effects ($\mu(n) = 0$, with indices omitted) are neglect. Under these conditions, upon applying the relations in Eqs. \eqref{continuidade}--\eqref{continuidade3} to Eq. \eqref{soliton}, while neglecting terms of order $\mathcal{O}(\Delta s^4)$, the resulting continuous-domain soliton is expressed as:


\begin{strip}
\begin{multline}
2i\,\dfrac{\partial \Phi}{\partial t}
-
\underbrace{
\varepsilon\left[
\pm{\Lambda^{3}}
\pm\,6{\Lambda^{2}}\Phi
- 2 \sigma\Phi
- 2 i {\zeta}_{\varepsilon}\Phi
\pm3\left|\Phi\right|^{2}\Phi
-\chi_l\frac{\Delta s^2}{2}\frac{\partial^2 \Phi}{\partial s^2}
\right]
}_{\text{Classical Damped Externally Driven Nonlinear Schrödinger Equation}}
\\
-
\underbrace{
\varepsilon^2\left[
\overline{\mathcal{L}}_{1} \Phi^{2}
+ \overline{\mathcal{L}}_2
+ \overline{\mathcal{L}}_3\left|\Phi\right|^{2}
+ \overline{\mathcal{L}}_{4}  \Phi
+
\overline{\mathcal{L}}_{5}\left|\Phi\right|^{2}\Phi
- \dfrac{15\left|\Phi\right|^{4}\Phi}{8}
\right.
}_{\text{Newly added terms}}
\\
\underbrace{
\left.
+\chi_l^2\frac{3\Delta s^2}{4}\frac{\partial^2 \Phi}{\partial s^2}
\pm 3\chi_l\Delta s^2\left[
\Lambda^2 \frac{\partial^2 \Phi}{\partial s^2}
- \frac{1}{4}\Phi^2\frac{\partial^2 \bar{\Phi}}{\partial s^2}
+ \frac{1}{4}\frac{\partial^2 \left[{\Phi}\left|{\Phi}\right|^2\right]}{\partial s^2}
+ \frac{1}{2}\left|{\Phi}\right|^2\frac{\partial^2 {\Phi}}{\partial s^2}
\right]
\right]
}_{\text{Newly added terms}}
=0,
\label{solitonCONTINUO}
\end{multline}
\end{strip}where $\Phi=\overline{A (s,t)}$.

Equation \eqref{solitonCONTINUO} is composed of terms classified according to the order of the perturbation factor $\varepsilon$. At first order, a classical formulation of the NLS is obtained, whose stability has been established for decades, both for bounded spatial domains \cite{terrones1990stability} and infinite spatial domains \cite{barashenkov1996existence}. However, when attention is directed to terms proportional to $\varepsilon^2$, the stability of the phenomenon becomes unreported, particularly due to the emergence of fifth-order nonlinearities of the form $\left|\Phi\right|^{4}\Phi$. One may note that, even when Eq. \eqref{solitonCONTINUO} is approximated at first order in $\varepsilon$, the parameters composing the equation remain coupled. The external excitation, described by the variable $\Lambda$, multiplies the function $\Phi$, and its effects are therefore balanced by the variable $\sigma$. Although physical parameter combinations capable of stabilizing the first-order equation in $\varepsilon$ theoretically exist, large values of $\Lambda$ compromise the predictive capability of the model, thus requiring the inclusion of the additional terms.

\begin{figure}[ht!]
    \centering
    \includegraphics[width=0.46\textwidth]{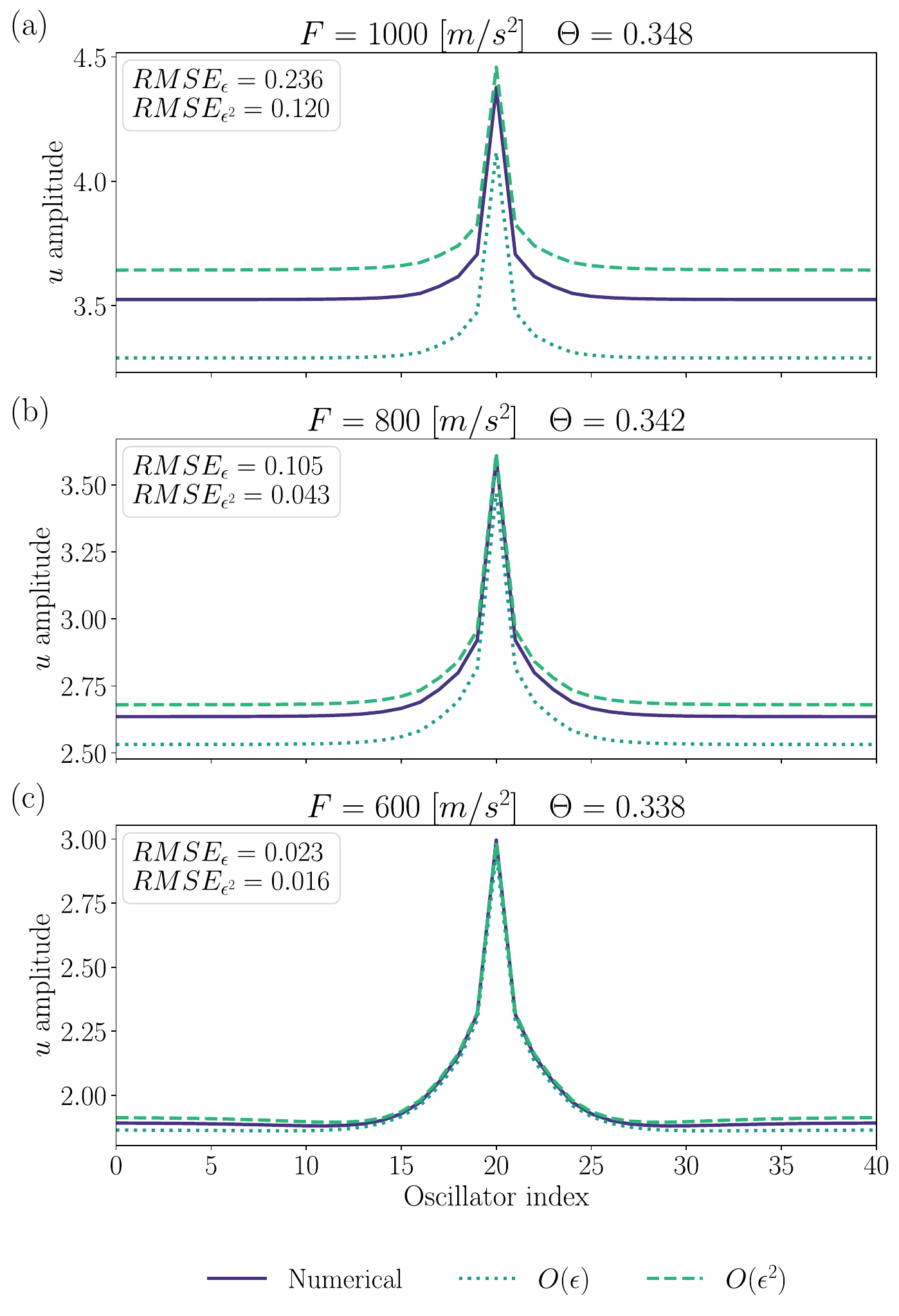}
    \caption{Comparison between analytical responses obtained using the method of multiple scales with a single time scale and with two time scales, and the numerical solution obtained from the integration of Eq.~\eqref{Eq mec normalizada} for the hardening case ($\pm = 1$). Panels a)--c) present the comparison for different excitation amplitudes. Numerical validation considers all oscillators initially at rest. Analytical solutions are obtained over 100 periods using the DOP853 method with tolerances $rtol=10^{-6}$ and $atol=10^{-8}$. Numerical integration is performed using the LSODA method with $rtol=10^{-6}$ and $atol=10^{-8}$ over $500$ cycles, with the first $200$ discarded as transient and $1000$ points adopted per cycle. The parameters are $\zeta=0.3/100$, $\omega_0=100$ [Hz], $\beta=2.5\times10^9\mathrm{[N/(kg\cdot m^3)]}$, and $\varepsilon=0.01$.
}
    \label{ValidationHardening}
\end{figure}

\begin{figure}[ht!]
    \centering
    \includegraphics[width=0.46\textwidth]{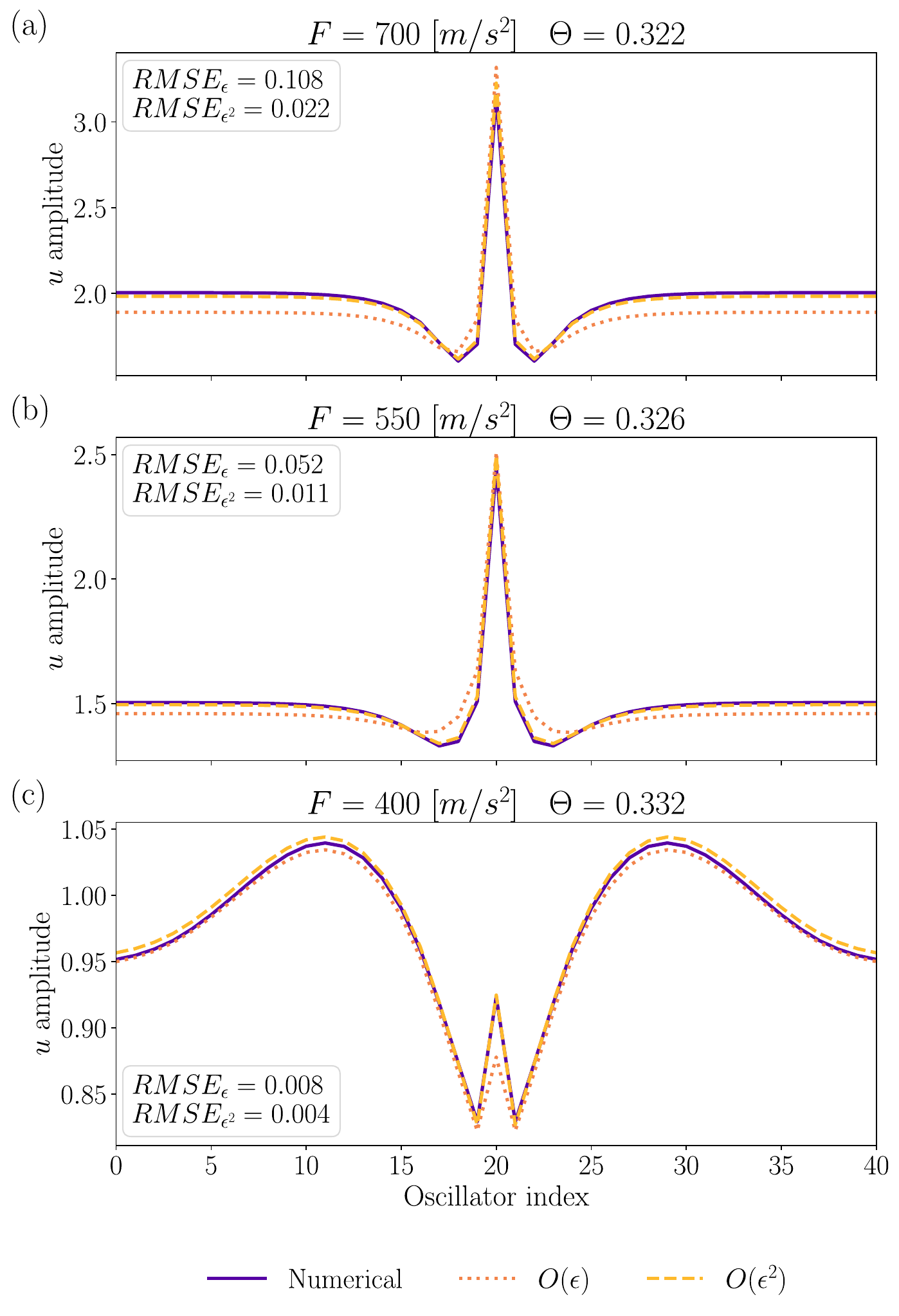}
    \caption{Comparison between analytical responses obtained using the method of multiple scales with a single time scale and with two time scales, and the numerical solution obtained from the integration of Eq.~\eqref{Eq mec normalizada} for the softening case ($\pm = -1$). Panels a)--c) present the comparison for different excitation amplitudes. Numerical validation considers all oscillators initially at rest. Analytical solutions are obtained over 100 periods using the DOP853 method with tolerances $rtol=10^{-6}$ and $atol=10^{-8}$. Numerical integration is performed using the LSODA method with $rtol=10^{-6}$ and $atol=10^{-8}$ over $500$ cycles, with the first $200$ discarded as transient and $1000$ points adopted per cycle. The parameters are $\zeta=0.3/100$, $\omega_0=100$ [Hz], $\beta=2.5\times10^9\mathrm{[N/(kg\cdot m^3)]}$, and $\varepsilon=0.01$.
}
    \label{ValidationSoftening}
\end{figure}

To investigate this premise, Figures \ref{ValidationHardening} and \ref{ValidationSoftening} are presented for the hardening and softening cases, respectively. The comparison is performed using the Root Mean Squared Error ($RMSE$) metric, computed for both the first-order perturbation approximation ($RMSE_{\varepsilon}$) and the second-order perturbation approximation ($RMSE_{\varepsilon^2}$). The results show that both $RMSE_{\varepsilon^2}$ and $RMSE_{\varepsilon}$ increase with excitation intensity for the hardening and softening regimes, indicating that the accuracy of the estimates improves under weaker excitation conditions. Moreover, $RMSE_{\varepsilon}$ exceeds $RMSE_{\varepsilon^2}$, demonstrating the greater sensitivity of this metric to changes in $F$.

It should be emphasized that the solitons shown in Figures \ref{ValidationHardening} and \ref{ValidationSoftening} are obtained for fixed values of the excitation frequency $\Theta$. Since the $RMSE$ varies across the frequency domain, the accuracy of the first-order approximation in $\varepsilon$ depends on the excitation condition. Consequently, although the approximation may accurately describe the lattice response at specific frequencies, its validity cannot be generalized to the entire excitation range. In this context, the numerical sweeps presented hereafter incorporate all terms of Eq. \eqref{soliton}, rather than being restricted to those associated with the classical NLS.

This section develops the theoretical features inherent to discrete solitons associated with superharmonic resonance. Through the method of multiple scales, the role of phase in the composition of the response is demonstrated, as the steady-state response involves multiple interacting frequencies. Although the NLS is associated with the first order of perturbation $\varepsilon$, the use of higher-order perturbations leads to relations whose stability has not been reported in the literature. Given the absence of closed-form solutions for Eq. \eqref{solitonCONTINUO}, the results of this study are obtained numerically through integration of Eq. \eqref{soliton}. Furthermore, owing to the difficulty of stabilizing the periodic case, disorder is introduced as a mechanism for soliton nucleation, in agreement with previous studies reporting an expansion of the stability diagram induced by impurities in nonlinear lattices \cite{alexeeva2000impurity,barbosa2024standing}.

\section{Duffing oscillator frequency response in superharmonic regime}
\label{Duffing oscillator frequency response in superharmonic regimes}

\begin{figure*}[ht!]
    \centering
    \includegraphics[width=1\textwidth]{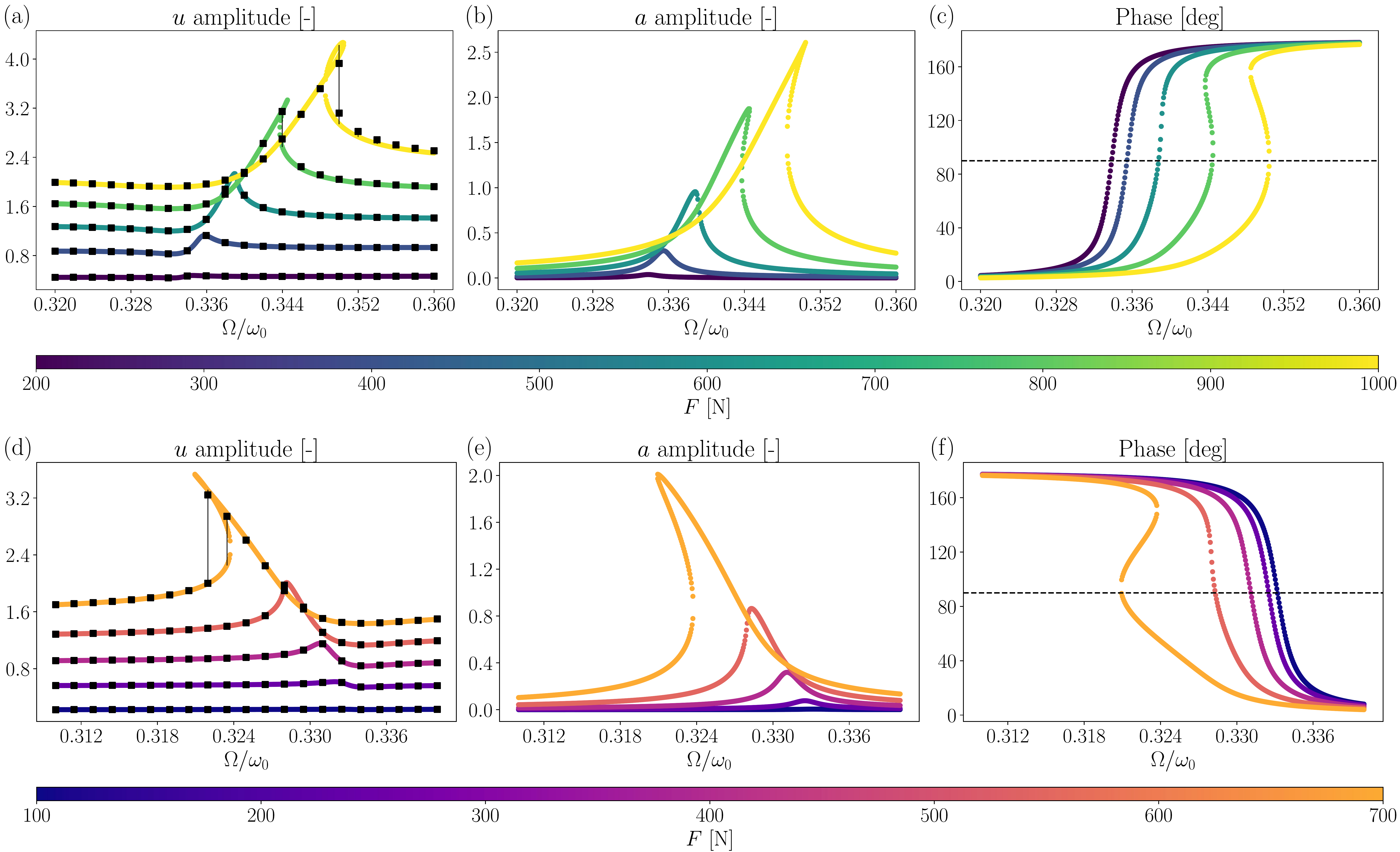}
    \caption{Simulation of the 1-DOF system illustrating the limitations of an isolated amplitude analysis. Panels a)--c) correspond to the hardening case, while panels d)--f) represent the softening condition. The three columns show the amplitude of $u_n$, the amplitude $a_n$, and the phase $\gamma_n$, respectively. Numerical validation is performed by selecting points on the $u_n$ analytical curve as initial conditions for the numerical integration of Eq.~\eqref{Eq mec normalizada}, producing the black squares. Differences between squares and analytical predictions indicate the limits of the multiple-scales approximation. Integration is performed with the LSODA method using $rtol=10^{-6}$ and $atol=10^{-8}$, simulating $500$ cycles (first $200$ discarded as transient) with $1000$ points per cycle. A total of $21$ validation points are used. Parameters: $\zeta=0.3/100$, $\omega_0=100$ [Hz], $\beta=2.5\times10^9\mathrm{[N/(kg\cdot m^3)]}$, and $\varepsilon=0.01$ .Excitation values for the softening regime are reduced to achieve numerical convergence.}
    \label{1DOF}
\end{figure*}
 
This section analyzes the dynamic behavior of the Duffing oscillator in the superharmonic regime, considered individually. The purpose of this study is to establish a basis for comparison between individual and collective dynamics, thereby enabling the identification of energy flow along the oscillators. The numerical sweeps are performed in both the hardening and softening regimes, under the assumptions of base acceleration and linear damping. It should be noted that alternative strategies for extracting the steady-state response of the single-degree of freedom 1-DOF system may be employed \cite{lenci2022exact}. However, the method of multiple scales is adopted to maintain consistency with the soliton derivation.
 
The mathematical description of 1-DOF system is obtained by suppressing the coupling and disorder terms in Eq. \eqref{Eq mec normalizada}, consequently yielding the decoupled form of the system. This formulation is equivalent to the response of an isolated oscillator, for which the following polar representation is adopted:
\begin{equation}
A_{n}(t) = \dfrac{a_{n}(t)\, e^{- i \gamma_{n}(t)}}{2},
\label{polar}
\end{equation}
with $a_{n}(t)$ and $\gamma_{n}(t)$ being the real variables associated with the amplitude and phase of $A_{n}(t)$, respectively. By introducing the polar representation into the restitution equation \eqref{soliton}, while separating the real and imaginary components and imposing the steady-state condition, two equations are obtained: 

\begin{equation}
\sin\!\left(\gamma_{n}\right)
=
\dfrac{Z X - V W - V Y + X a_{n}\,\varepsilon\,\sigma}
     {W^{2} + X^{2} - Y^{2}},
\label{seno_fase}     
\end{equation} and

\begin{equation}
\cos\!\left(\gamma_{n}\right)
=
\dfrac{- V X - \left(Z + a_{n}\,\varepsilon\,\sigma\right)\left(W - Y\right)}
     {W^{2} + X^{2} - Y^{2}},
\label{coseno_fase}         
\end{equation}where the variables $Z$, $X$, $V$, $W$, and $Y$ are defined in Appendix \ref{Auxiliary variables employed in the mathematical formulation}, Table \ref{auxiliares_sen_coseno}. From the algebraic manipulation of Equations \eqref{seno_fase} and \eqref{coseno_fase}\footnote{$\sin^{2}\left(\gamma_{n}\right) + \cos^{2}\left(\gamma_{n}\right) = 1$}, an algebraic equation for the amplitude $a_n$ of the 1-DOF system is obtained and can be solved. This amplitude does not correspond to the oscillator displacement, but rather to the components associated with $3\Theta$. The analytical solution for $u_n(t)$ follows from substituting Equations \eqref{solu_u0} and \eqref{solu_u01} into \eqref{solução}, together with the polar representation of Eq.\eqref{polar}, yielding\footnote{This equation holds for both the 1-DOF and N-DOF cases, with $\mu(n) = 0$ in the 1-DOF case.}:

{\small
\begin{multline}
u_{n}(t) = a_{n} \cos\!\big(3t\Theta - \gamma_{n}\big) + 2\Lambda_n \cos\!\big(t\Theta\big) \\
+ \varepsilon\bigg(\bigg(\dfrac{2 \Lambda_n \mu(n)}{\Theta^{2} - 1}
\pm\dfrac{3\Lambda_n a_{n}^{2}}{\Theta^{2} - 1} \pm\dfrac{3\Lambda_n^2 a_{n}\cos\!\big(\gamma_{n}\big)}{4\Theta(\Theta - 1)} \\
\pm \dfrac{6\Lambda_n^{3}}{\Theta^{2} - 1}\bigg) \cos\!\big(t\Theta\big) 
+ \bigg(\pm\dfrac{3\Lambda_n^{2}a_{n}\sin\!\big(\gamma_{n}\big)}{4\Theta(\Theta - 1)} \\ 
- \dfrac{4\Theta\Lambda_n{\zeta_n}_{\varepsilon}}{\Theta^{2} - 1}\bigg) \sin\!\big(t\Theta\big)\bigg) + \text{NRT},
\label{solu_1DOF}
\end{multline}
}where NRT refers to the non-resonant terms, which are neglected in the numerical simulations.


A direct consequence of the superposition of cosine and sine terms with different frequencies is the requirement to evaluate the steady-state amplitude based on the individual contributions of its components. In other words, the amplitude associated with $u_{n}$ does not coincide with that obtained from the secular terms presented in Eq. \eqref{soliton}, since these describe the variable $A_n$, whose amplitude is defined by $a_n$ in Eq. \eqref{polar}. To illustrate this issue, the simulation presented in Fig. \ref{1DOF} is examined. Panels a), b), and c) display the hardening behavior, whereas panels d), e), and f) correspond to the softening condition. The figure is organized into three columns: the first presents the amplitude of $u_n$, the second the amplitude of $A_n$ ($a_n$), and the third the phase of $A_n$ ($\gamma$). The results highlight the distinct shapes of the amplitude curves while preserving a comparable order of magnitude, thereby reinforcing the inadequacy of an isolated analysis. As the analytical formulation is derived using the method of multiple scales, a numerical validation is superimposed on the amplitude curves of $u_n$ (black square markers),  confirming the validity of the solution in Eq. \eqref{solu_1DOF}.

In conclusion, this section establishes a reference for analyzing energy flow in oscillators subjected to collective dynamics. With the 1-DOF behavior characterized, abrupt increases in oscillator amplitude can be interpreted in light of the discrete soliton equation. In contrast to the commonly reported behavior under primary excitation, the oscillator phase plays a critical role in determining the amplitude of $u_n$, a result supported by numerical validation.

\begin{figure*}[ht!]
    \centering
    \includegraphics[width=1\textwidth]{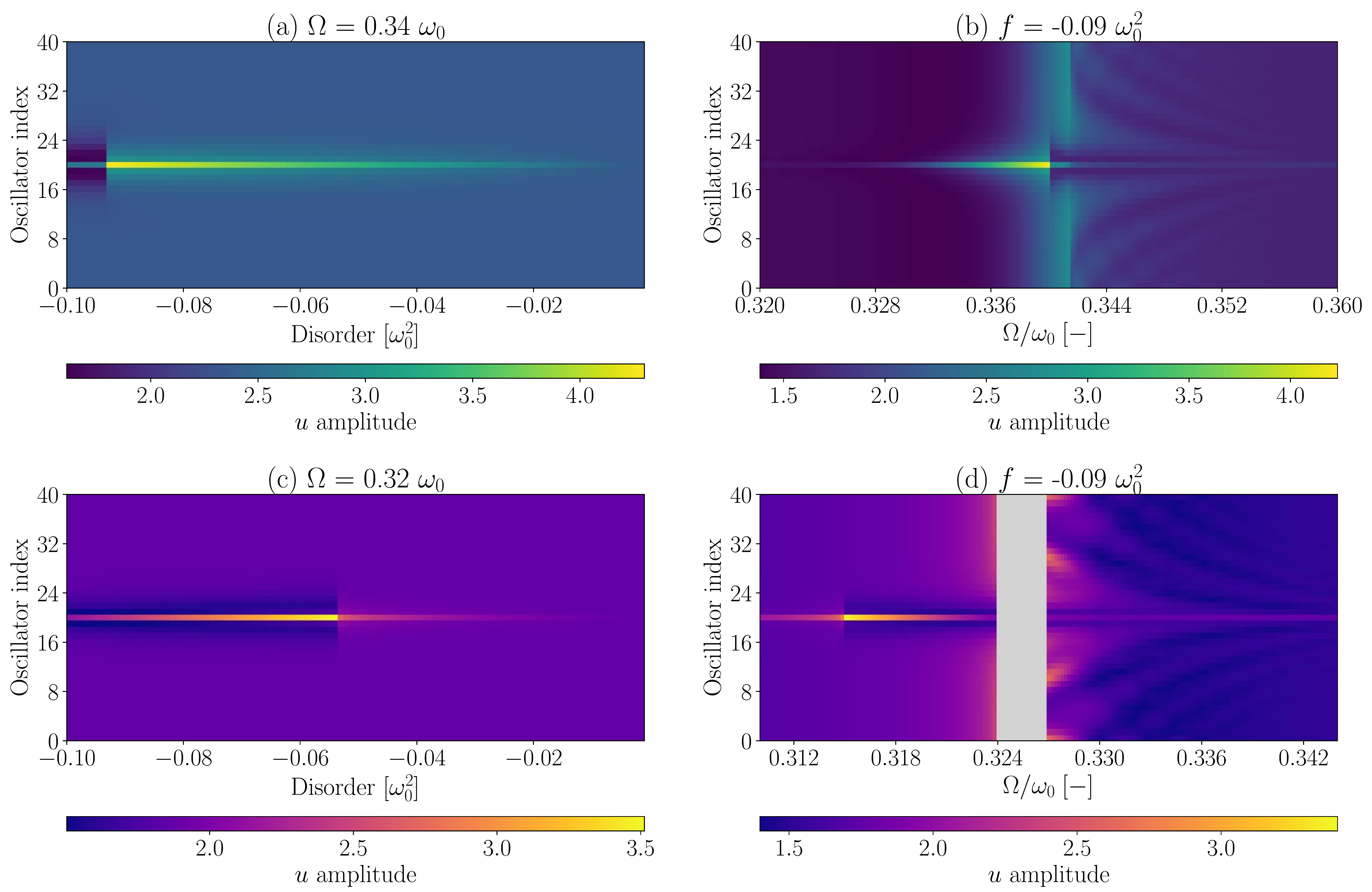}
    \caption{Numerical sweeps of soliton nucleation for hardening (a–b) and softening (c–d) regimes. Panels (a,c): fixed excitation frequency with varying disorder; (b,d): fixed disorder with varying frequency. Simulations use the DOP853 method ($rtol=10^{-6}$, $atol=10^{-8}$) over 100 periods. Parameters: $\zeta=0.3/100$, $\omega_0=100$ Hz, $\beta=2.5\times10^9 [\mathrm{N/(kg\cdot m^3)}]$, $e=0.01$, and $F=700\mathrm{m/s^2}$. Gray region (c–d): no steady-state convergence. Absence of nonlinear coupling. Linear coupling up to third order ($\alpha=3$), with $C_l^{(1)} = 2.5\%,\omega_0^2$, $C_l^{(2)} = C_l^{(1)}/4$, and $C_l^{(3)} = C_l^{(1)}/9$. Periodic parameters, except for the introduced disorder. Numerical continuation imposed on the frequency domain.}
    \label{Prova_de_soliton}
\end{figure*}

\section{Energy nucleation and stationary soliton analysis}
\label{Energy nucleation and stationary soliton analysis}

This section demonstrates the nucleation of solitons from a set of oscillators. The primary goal is to establish the phenomenon numerically, focusing on three aspects: the influence of disorder, the number of oscillators, and the robustness of the phenomenon against uncertainties in the physical parameters. The analyses are conducted in terms of the amplitudes of $u_n$ (or $U_n$ when considering lattice dimensions), since the solution for $A_n$ does not fully characterize the steady-state amplitude of the metastructure. It should be noted that Eq. \eqref{restituição} reflects a stationary soliton only when the temporal variation of $A_n$ vanishes. Accordingly, a convergence metric is defined in the present work as follows:

\begin{equation}
\max_{n} \left| \dfrac{d A_n}{dt} \right| < 10^{-7}
\;\;\text{and}\;\;
\max_{n} \left| \dfrac{d^2 A_n}{dt^2} \right| < 10^{-8}.
\label{critério de regime permanente}
\end{equation} It is also emphasized that the simulations may be conducted from motionless oscillators, treated independently, i.e., without numerical continuation, conditions explicitly stated over the figures.

The frequency-domain simulations are derived from time-domain integrations. Specifically, to obtain an amplitude profile at a prescribed value of $\Theta$, the governing equations are integrated until the system satisfies Eq. \eqref{critério de regime permanente}; otherwise, the solution is classified as non-convergent. Fig. \ref{Timeevo} in Appendix \ref{Supplementary numerical simulations} illustrates this procedure by presenting spatio-temporal simulations of Eq. \eqref{soliton}.

\begin{figure*}[ht!]
    \centering
    \includegraphics[width=1\textwidth]{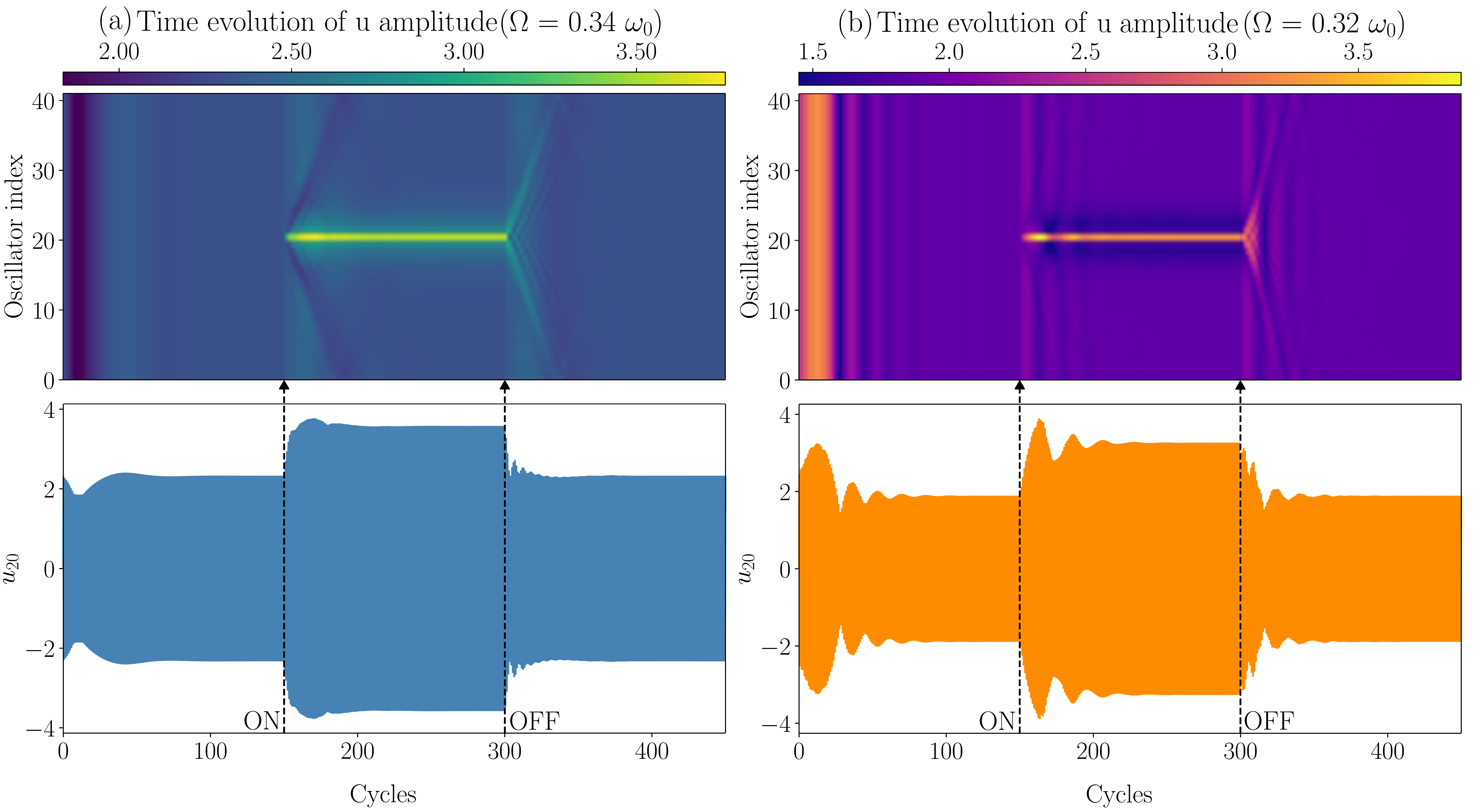}
    \caption{Simulations with disorder toggled on/off for hardening (a) and softening (b) regimes, showing induced localization and subsequent delocalization upon removal. The temporal response of the perturbed oscillator is shown below each heat map, indicating non-sustained localization. Integration of Eq.~\eqref{Eq mec normalizada} using the LSODA algorithm ($rtol=10^{-6}$, $atol=10^{-8}$). Parameters: $\zeta=0.003$, $\omega_0=100$ Hz, $\beta=2.5\times10^9\,\mathrm{N/(kg\cdot m^3)}$, $e=0.01$, $F=700\,\mathrm{m/s^2}$, and disorder $f=-0.06k_l$. Absence of nonlinear coupling. Linear coupling up to third order ($\alpha=3$), with $C_l^{(1)} = 2.5\% \,\omega_0^2$, $C_l^{(2)} = C_l^{(1)}/4$, and $C_l^{(3)} = C_l^{(1)}/9$. Periodic parameters, except for the introduced disorder. Numerical continuation imposed in time domain.}
    \label{TimeResponse}
\end{figure*}

\subsection{On the influence of the disorder}
\label{On the influence of the disorder}

It is established in the literature that energy localization in chains of Duffing oscillators can occur in periodic structures with spatially uniform physical properties. Concurrently, disorder is known to enhance the stability of the phenomenon and promote spontaneous nucleation, particularly through the breaking of periodicity toward decreasing natural frequencies of the oscillators. In this context, the verification of such a strategy in superharmonic regimes constitutes a plausible hypothesis, which is investigated in this subsection. Fig. \eqref{Prova_de_soliton} presents four numerical sweeps that elucidate such idea: two corresponding to the hardening case (a) and (b), and two to the softening case (c) and (d). Figures (a) and (c) maintain a constant excitation frequency while assessing the effect of impurity on soliton nucleation, whereas Figures (b) and (d) perform the complementary analysis by fixing the disorder and varying the excitation frequency.

An immediate conclusion can be drawn from Fig. \ref{Prova_de_soliton}: energy nucleation can be tuned through an appropriate selection of disorder, which depends on the operating frequency (or bandwidth) of the metastructure. A second conclusion pertains to the nature of the nonlinearity: not only does the optimal level of disorder vary depending on whether a hardening or softening system is considered, but the mechanisms underlying abrupt nucleation also differ in each case. For simulations involving negative nonlinearity, a gray rectangle is delineated by a dashed line. This region does not satisfy the criterion defined in Relations \eqref{critério de regime permanente}, indicating that $A_n$ not converge to a stable profile. Such lack of convergence did not necessarily imply chaotic dynamics (simulation \ref{Timeevo} (e) - (f) in Appendix \ref{Supplementary numerical simulations}), but rather the absence of a temporally stable waveform, which may still exhibit periodic behavior (simulation \ref{Timeevo} (c) - (d) in Appendix \ref{Supplementary numerical simulations}).

By revisiting the secular terms associated with the first power of $\varepsilon$ in Eq. \eqref{Seculares_1}, the existence of localized nonlinear modes in periodic chains is readily inferred, as the NLS emerges at this order. However, in contrast to the primary resonance case, the soliton parameters are intrinsically correlated. The term equivalent to the external excitation ($(\pm){\Lambda^{3}{n}} e^{i T{1}\sigma}$) depends on $\Lambda$, which also multiplies $A_n$. This observation becomes more pronounced in higher-order secular terms in Eq. \eqref{Seculares_2}, where the physical parameters of the lattice exhibit nonlinear coupling, hindering the identification of their isolated effects, as performed in \cite{BARBOSA2024111358} for the fundamental resonance.

\begin{figure*}[ht!]
    \centering
    \includegraphics[width=1\textwidth]{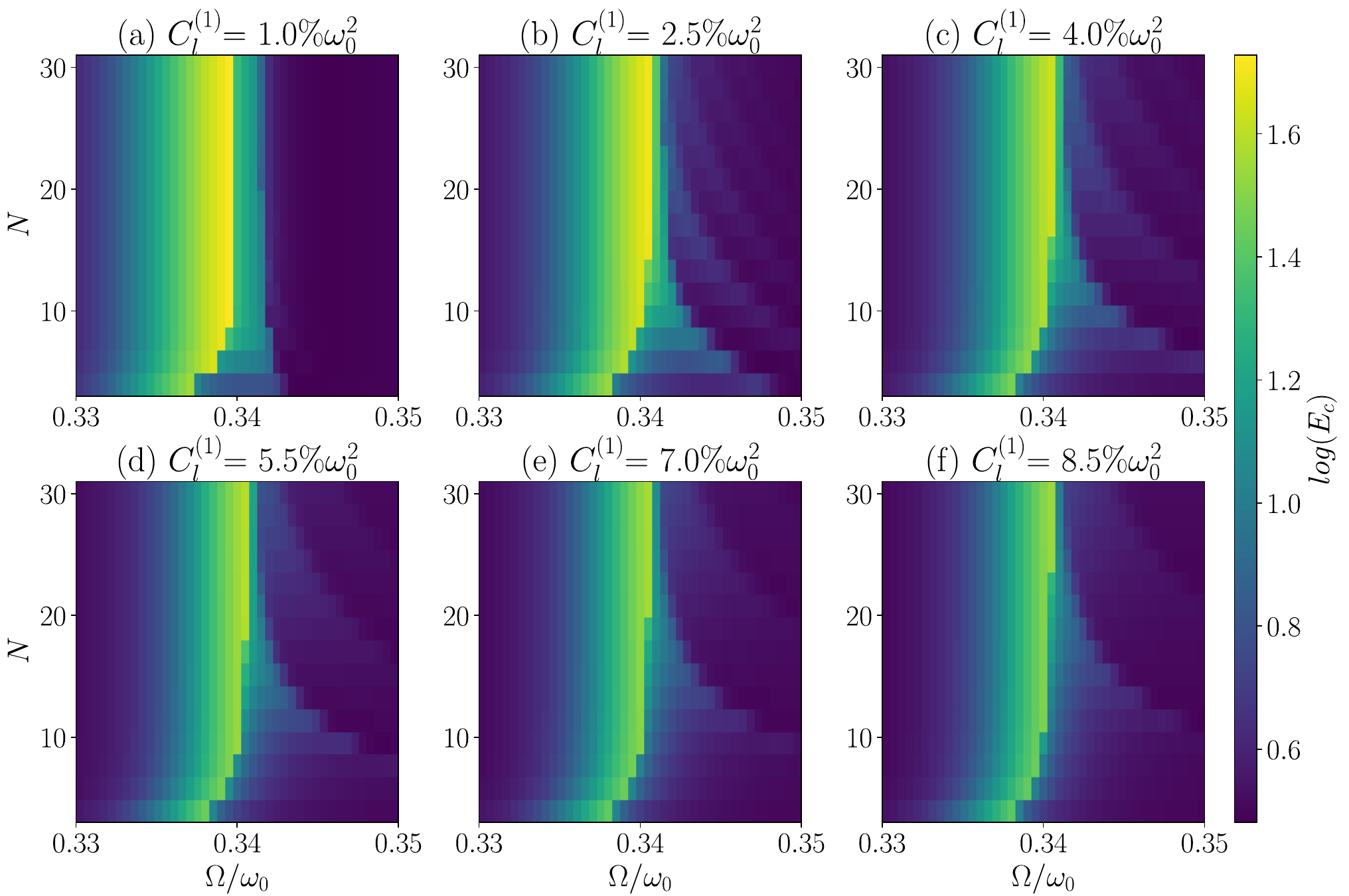}
    \caption{Numerical sweep conducted to determine the saturation values of the number of oscillators in the hardening regime ($\pm = 1$). Simulations are performed using the DOP853 method with tolerances $rtol = 10^{-6}$ and $atol = 10^{-8}$. The adopted parameters are $\zeta = 0.3/100$, $\omega_0 = 100\,\mathrm{Hz}$, $\beta = 2.5 \times 10^9\,\mathrm{N/(kg \cdot m^3)}$, $e = 0.01$, and $F = 700\,\mathrm{m/s^2}$. Nonlinear coupling is neglected, while linear coupling is considered up to third order ($\alpha = 3$), with coefficients $C_l^{(2)} = C_l^{(1)}/4$ and $C_l^{(3)} = C_l^{(1)}/9$. All parameters are periodic except for the introduced disorder ($f = -0.05\omega_0^2$). Numerical continuation is not employed.}
    \label{HardeningFig5}
\end{figure*}

From such standpoint, Fig. \ref{TimeResponse} presents two sets of simulations, (a) corresponding to the hardening regime and (b) to the softening regime, in which system disorder is selectively activated and deactivated over the course of the simulation. In both cases, the introduction of disorder into the system leads to energy localization, whereas its removal results in delocalization. For clarity, the temporal response of the oscillator at which disorder is introduced is displayed below the heat map. As observed, the resulting solution is not sustained over time. No set of physical parameters yielding localized solutions in the absence of disorder was identified, although this does not imply that such solutions are unattainable. Another relevant design consideration is that, analogously to the primary case, only disorder that reduces the natural frequency was numerically observed to induce soliton nucleation, implying that positive values of $f(n)$ do not promote energy localization (see Fig. \ref{prova_de_solitonIMPpos} in Appendix \ref{Supplementary numerical simulations}).

\subsection{On the saturation of the number of oscillators}
\label{On the saturation of the number of oscillators}

\begin{figure*}[ht!]
    \centering
    \includegraphics[width=1\textwidth]{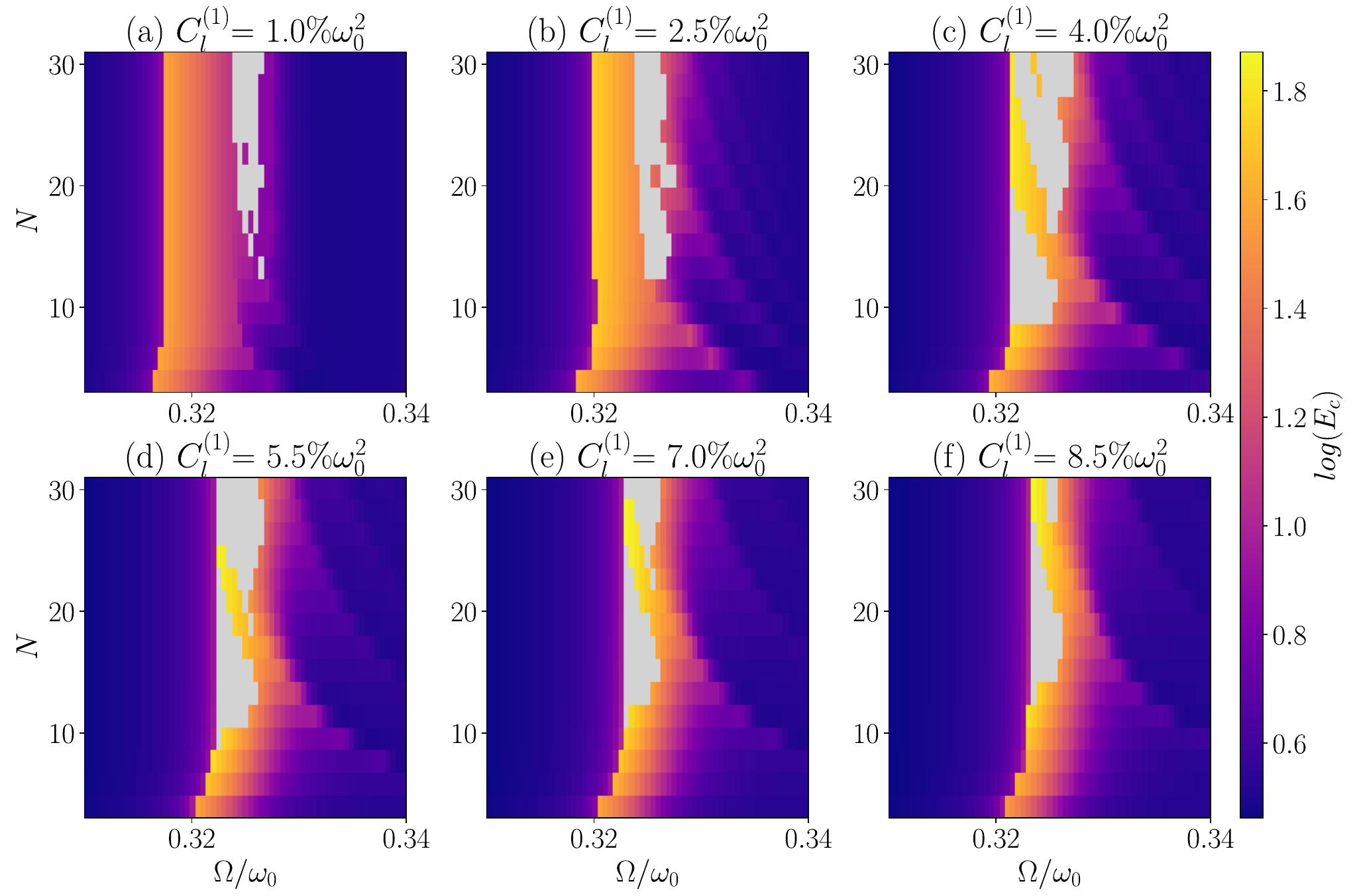}
    \caption{Numerical sweep conducted to determine the saturation values of the number of oscillators in the softening regime ($\pm = -1$). Simulations are performed using the DOP853 method with tolerances $rtol = 10^{-6}$ and $atol = 10^{-8}$. The adopted parameters are $\zeta = 0.3/100$, $\omega_0 = 100\,\mathrm{Hz}$, $\beta = 2.5 \times 10^9\,\mathrm{N/(kg \cdot m^3)}$, $e = 0.01$, and $F = 700\,\mathrm{m/s^2}$. Nonlinear coupling is neglected, while linear coupling is considered up to third order ($\alpha = 3$), with coefficients $C_l^{(2)} = C_l^{(1)}/4$ and $C_l^{(3)} = C_l^{(1)}/9$. All parameters are periodic except for the introduced disorder ($f = -0.05\omega_0^2$). Numerical continuation is not employed.}
    \label{SofteningFig5}
\end{figure*}

From the metastructure design perspective, an appropriate selection of the number of DOF is critical, due to the trade-off between the minimum number of oscillators required for the phenomenon and the maximization of the number of oscillators whose energy migrates toward the soliton core. It should be noted that the admissible range of $N$ is established after fixing the physical parameters of the lattice. As an illustration of this issue, the selection of the number of oscillators can be addressed from the continuum hypothesis stand point (limit \eqref{continuidade}). Consistent with the spatial adimensionalization adopted in \cite{BARBOSA2024111358,barbosa2024standing,partI,partII}, the allocation of oscillators within the spatial domain governed by the Schrödinger equation depends on the coupling stiffness. In other words, the number of oscillators required for the manifestation of the phenomenon is determined by $k_c$, a relationship that extends to the perturbation parameter, for instance.

The current subsection investigates the influence of a gradual increase in $N$ as the coupling strength varies. In order to facilitate the analysis, a set of simplifying assumptions is introduced, slightly restricting the generality of Eq. \eqref{Equação mecânica dim}. In particular, nonlinear couplings are neglected and the coupling order is limited to $\alpha = 3$, with $C_l^{(2)} = C_l^{(1)}/4$ and $C_l^{(3)} = C_l^{(1)}/9$. In Appendix \ref{Supplementary numerical simulations} (Fig. \ref{ParamKnlNH} for hardening case and Fig. \ref{ParamKnlNS} for the softening one), specific cases incorporating nonlinear couplings are presented; however, a detailed analysis of their effects lies beyond the scope of this study.

To quantify the system response, we evaluate the kinetic energy $E_c$ of the oscillator where the disorder is introduced over a full period of motion:

{\small
\begin{multline}
E_c = \int_{0}^{\dfrac{2\pi}{\Theta}} \left(\dfrac{d}{dt} u_n(t)\right)^2 \, dt
= \\ \pi \Theta \left( \Pi_1^2 \varepsilon^2 + \Pi_2^2 \varepsilon^2 + 4 \Pi_2 \Lambda_n \varepsilon + 9 a_n^2 + 4 \Lambda_n^2 \right),
\label{E_c}
\end{multline}
}where $\Pi_1$ and $\Pi_2$ are defined in Table \ref{tabela Ec}. Figures \ref{HardeningFig5} (hardening regime) and \ref{SofteningFig5} (softening regime) present simulation panels evaluating the amplitude of $\log(E_c)$ as a function of excitation frequency and the number of oscillators for different coupling strengths. In both regimes, the saturation of $N$ depends on the excitation frequency and on the selected coupling strength. Furthermore, an unbounded increase in the number of oscillators leads to a reduction in the kinetic energy for specific frequency ranges, consistent with the physical expectation that localized modes exhibit preferential formation frequencies. The non-convergence issue (relations \eqref{critério de regime permanente}) for negative nonlinearity values is also governed by the number of oscillators, becoming more pronounced as $N$ increases. Additionally, increasing $C_l^{(1)}$ raises the number of oscillators required to achieve energy saturation, with this effect being more readily observed in the hardening regime. Another relevant aspect concerns the simulation condition, namely the absence of numerical continuation. In Figures \ref{Prova_de_soliton} and \ref{TimeResponse}, the solutions of Eq. \eqref{soliton} are initialized from sufficiently close prior states (either in frequency or time), ensuring convergence within neighboring basins of attraction. In contrast, Figures \ref{HardeningFig5} and \ref{SofteningFig5} are obtained from initially motionless oscillators, implying spontaneous nucleation; a similar mechanism observed for solitons emerging from primary excitation frequencies. It should be emphasized that, despite the conclusions drawn from simulations represented in Fig. \ref{Prova_de_soliton} and Fig. \ref{TimeResponse}, alternative coupling configurations may yield different outcomes, particularly under the assumption of numerical continuity.

\subsection{On the robustness of the phenomenon to uncertainties in physical parameters}
\label{On the robustness of the phenomenon to uncertainties in physical parameters}

\begin{figure*}[ht!]
    \centering
    \includegraphics[width=1\textwidth]{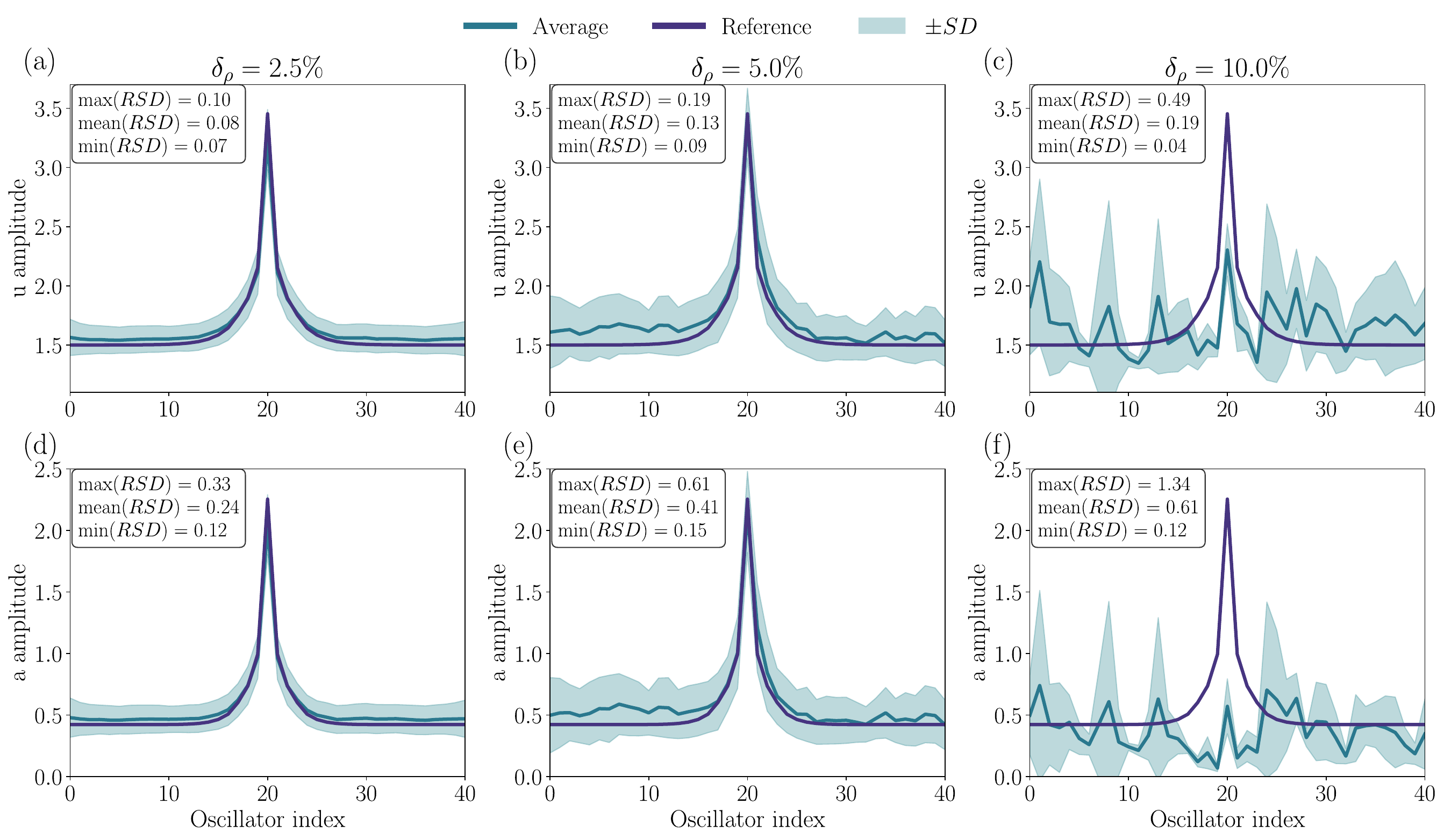}
    \caption{Simulations under nondeterministic regimes for the hardening case ($\pm = 1$). The simulations are performed using the DOP853 method with tolerances $rtol = 10^{-6}$ and $atol = 10^{-8}$. The adopted nominal parameters are $\zeta = 0.3/100$, $\omega_0 = 100\,\mathrm{Hz}$, $\beta = 2.5 \times 10^9\,\mathrm{N/(kg \cdot m^3)}$, $e = 0.01$, $F = 700\,\mathrm{m/s^2}$ and $\Theta =0.336 $. Nonlinear coupling is neglected, while linear coupling is considered up to third order ($\alpha = 3$), with $C_l^{(1)} = 2.5\% \, \omega_0^2$, $C_l^{(2)} = C_l^{(1)}/4$, and $C_l^{(3)} = C_l^{(1)}/9$. The damping, natural frequency, and oscillator excitation are subjected to arbitrary variations described by Eq.\eqref{incertezas}, whereas the remaining parameters are kept periodic. Variations in the natural frequency are associated with the introduced disorder, with an intended disorder imposed on the central oscillator ($\tilde{f}(20) = -0.05\omega_0^2$, and $\tilde{f}(n) = 0$ for $n\neq20$). All solutions are obtained from rest initial conditions.}
    \label{PerturbedHardening}
\end{figure*}

\begin{figure*}[ht!]
    \centering
    \includegraphics[width=1\textwidth]{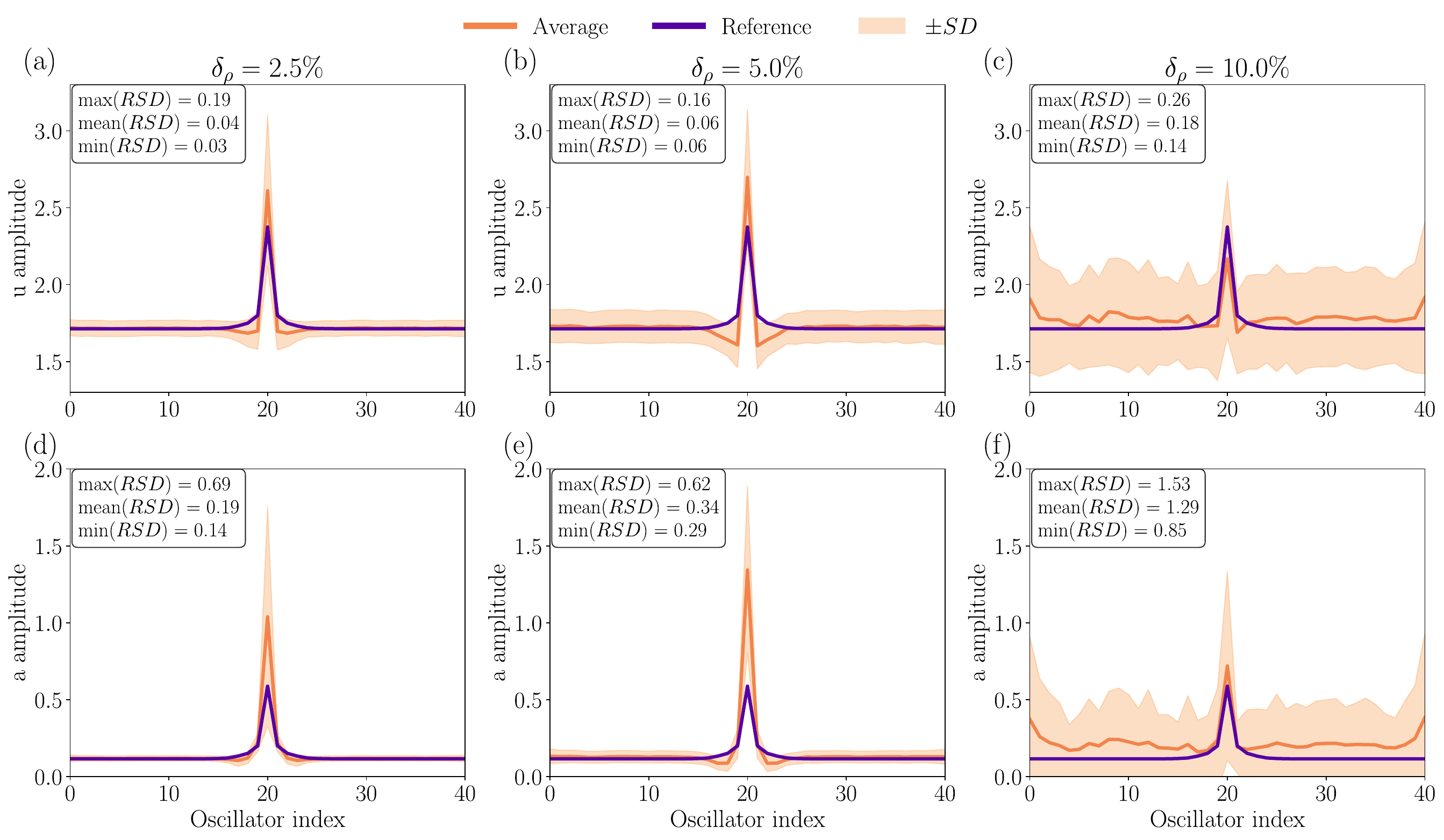}
    \caption{Simulations under nondeterministic regimes for the softening case ($\pm = -1$). The simulations are performed using the DOP853 method with tolerances $rtol = 10^{-6}$ and $atol = 10^{-8}$. The adopted nominal parameters are $\zeta = 0.3/100$, $\omega_0 = 100\,\mathrm{Hz}$, $\beta = 2.5 \times 10^9\,\mathrm{N/(kg \cdot m^3)}$, $e = 0.01$, $F = 700\,\mathrm{m/s^2}$, and $\Theta =0.312 $. Nonlinear coupling is neglected, while linear coupling is considered up to third order ($\alpha = 3$), with $C_l^{(1)} = 2.5\% \, \omega_0^2$, $C_l^{(2)} = C_l^{(1)}/4$, and $C_l^{(3)} = C_l^{(1)}/9$. The damping, natural frequency, and oscillator excitation are subjected to arbitrary variations described by Eq.\eqref{incertezas}, whereas the remaining parameters are kept periodic. Variations in the natural frequency are associated with the introduced disorder, with an intended disorder imposed on the central oscillator ($\tilde{f}(20) = -0.05\omega_0^2$, and $\tilde{f}(n) = 0$ for $n\neq20$). All solutions are obtained from rest initial conditions.}
    \label{PerturbedSoftening}
\end{figure*}

One of the defining characteristics of solitons concerns their stability against waveform dispersion. In fact, when compared with linear defect-induced localizations (Anderson localization, for instance), disorder-nucleated solitons exhibit broader tolerance ranges to uncertainties in the physical parameters \cite{BARBOSA2025134612}, thereby enabling an easier experimental implementation of the phenomenon \cite{partII}. In this context, the problem of superharmonic nucleation in lattices for which the determinism hypothesis no longer holds naturally emerges.

To address this problem, let us consider that, for an arbitrary variable $\rho(n)$, possible variations around its nominal value $\tilde{\rho}$ are described by:
\begin{equation}
\rho(n) = \tilde{\rho} + \delta_{\rho}\,\mathcal{U}(n),
\label{incertezas}
\end{equation}
where $\delta_{\rho}$ denotes the maximum admissible percentage variation of $\tilde{\rho} $, and $\mathcal{U}(n)$ represents the probability function associated with the spatial occurrence. For the sake of generality, it is assumed to follow a uniform distribution over the interval $[-1,1]$. In accordance with the lattice description shown in Fig. \ref{Sistema}, the parameters subjected to \eqref{incertezas} correspond to the damping, natural frequency (by the disorder function), and oscillator excitation, which are evaluated simultaneously.

Figures \ref{PerturbedHardening} and \ref{PerturbedSoftening} present simulations under nondeterministic conditions imposed on Eq. \eqref{soliton} for the hardening and softening cases, respectively. The simulations are conducted under three distinct uncertainty levels, imposed according to the probabilistic law defined by $\mathcal{U}(n)$. For each value of $\delta_{\rho}$, $1000$ samples are generated, and the corresponding numerical integrations are classified as solitons if they satisfied the convergence defined in \eqref{critério de regime permanente}, while preserving the energy localization.  Once Eq. \eqref{soliton} is simulated in the absence of impurities, a reference pattern for the spatial energy distribution can be established. More specifically, to assess the persistence of the phenomenon, the criteria employed in \cite{BARBOSA2025134612} are adopted. To determine whether the solution remains bounded, the oscillation at $n=n_i$ is considered localized when its amplitude satisfies:
\begin{equation}
0.25\,{\mathcal{R}\mathcal{E}\mathcal{F}} \leq |A(n_i,t)| \leq 4\, {\mathcal{R}\mathcal{E}\mathcal{F}},
\end{equation}
where ${\mathcal{R}\mathcal{E}\mathcal{F}}$ denotes the maximum amplitude obtained in the absence of uncertainties. In addition to the amplitude criterion, a spatial criterion is employed. After identifying the peaks of $|A(n,t)|$ (local maxima of the profile), the solution is classified as a soliton when the ratio between the second-highest peak and the highest peak does not exceed $0.5$. Despite the practicality of this approach, some considerations are necessary. The verification procedure is based on $A(n,t)$ rather than $u(n,t)$, unlike in \cite{BARBOSA2025134612}, once the problem addressed therein concerned nucleation around the primary harmonic. A further distinction lies in the nature of the nonlinearity, which in the present study includes both hardening and softening regimes.

For comparison purposes, Figures \ref{PerturbedHardening} and \ref{PerturbedSoftening} present the reference curve (without uncertainties), the mean response obtained from the simulations, and a shaded envelope bounded by the standard deviation (SD). As a measure of robustness, the Relative Standard Deviation (RSD) is adopted, defined as the ratio between the standard deviation of the amplitudes and their corresponding mean values. Given the system's multiple degrees of freedom, the maximum, minimum, and mean RSD values are reported in the figures. In view of the stochastic nature of the analysis, the percentage of convergent solutions is also required to support conclusions drawn from the figures (see Table \ref{tab:acceptance_rates}).

\begin{table}[htbp]
\centering
\caption{Percentage of solutions for which soliton dynamics remained preserved under different levels of uncertainty $\delta_{\rho}$.}
\label{tab:acceptance_rates}
\begin{tabular}{ccc}
\hline
\textbf{$\delta_{\rho}$ (\%)} & Hardening Case (\%) & Softening Case (\%) \\
\hline
1.0 & 97.1 & 100.0 \\
2.5  & 63.1 & 95.3 \\
5.0  & 4.0  & 56.8 \\
10.0 & 0.4  & 25.9 \\
\hline
\end{tabular}
\end{table}

In line with the simulations presented in this subsection, some conclusions can be drawn. First, regarding the persistence of the phenomenon, increasing $\delta_{\rho}$ reduces the number of solutions exhibiting soliton dynamics, with a more pronounced decline in the hardening regime than in the softening one. Furthermore, increasing $\delta_{\rho}$ also leads to an increase in the RSD for both regimes. For $u$ amplitude (Figures (a), (b) and (c)), a gradual increase in dispersion is observed for both cases, with comparable values at the highest uncertainty levels. In contrast, the $a$ amplitude (Figures (d), (e) and (f)) exhibits a distinct trend: although the hardening case displays an approximately monotonic increase in variability, the softening case undergoes a stronger increase at the highest value of $\delta_{\rho}$, reaching a relative standard deviation more than twice that observed for the hardening regime. Such findings indicate that softening systems are more sensitive to elevated levels of uncertainty, particularly with respect to the $A(n,t)$.

\section{Concluding remarks}
\label{Concluding remarks}

This article demonstrates the existence and controllability of stationary solitons in nonlinear metastructures under superharmonic resonance. The system under investigation, consisting of a chain of coupled Duffing oscillators, is subjected to disorder, therefore enabling extension of the modeling to aperiodic configurations. By employing the method of multiple scales, a class of Discrete Nonlinear Schrödinger-type Equation is derived. Owing to the use of higher-order multiple-scales expansions, the resulting solution captures nonlinear interactions not previously reported for multi-degree-of-freedom mechanical systems, which are mainly analyzed in the vicinity of primary resonance.

The analyses are performed from two complementary perspectives: (i) examination of the analytically derived expressions and (ii) their validation via numerical integration of the lattice equation. Such approach enables parallels between primary and superharmonic mechanical solitons. Notably, the phase of the envelope function constitutes a key parameter. Unlike the primary resonance case, the oscillator response is expressed in terms of sine and cosine components whose direct algebraic superposition is not possible without proper phase consideration, hence precluding a direct mapping between the soliton variable and the measurable displacement field, as demonstrated in the 1-DOF simulations. Additionally, the selected excitation mechanism introduces a further nuance: the coupling between external forcing and spatial disorder, mediated by the $\Lambda_n$ term, induces a dependence between forcing and localization mechanisms that is absent in primary resonance scenarios. More specifically, because $\Lambda_n$ depends on both the excitation amplitude $K_n$ and the excitation frequency $\Theta$, it becomes directly involved in both the forcing and localization mechanisms. In primary resonance, the influence of forcing and disorder can generally be decoupled, enabling an immediate interpretation of the corresponding stability diagrams. In the present case, however, the resulting cross-coupling introduces additional complexity, since modifications in the excitation conditions simultaneously alter the forcing contribution and the localization conditions.

Regarding the numerical findings on the role of disorder, some conclusions are established. First, the hypothesis of spontaneous nucleation induced by a reduction in the natural frequency of the components is confirmed for both hardening and softening regimes. Second, analogously to primary resonance phenomena, such nucleation exhibits strong sensitivity to excitation frequency, coupling strength, and the number of oscillators, as a result hindering the formulation of a generalized stability diagram. Finally, it is demonstrated that periodicity-breaking mechanisms that increase the natural frequencies of the oscillators do not support soliton formation, instead promoting energy delocalization.

From an engineering perspective, operating at submultiples of the fundamental frequency mitigates geometric scaling limitations. In particular, exploiting superharmonic resonances enables access to similar high-frequency dynamics under lower-frequency excitation, despite uncertainties in the physical parameters. For $\delta_{\rho}=2.5\%$, the soliton nucleation rate reaches $95.3\%$ in the softening regime and $63.1\%$ in the hardening regime. Although increasing uncertainty progressively reduces these rates and increases the RSD, localized states persist even for $\delta_{\rho}=10\%$. Overall, the softening regime exhibits greater persistence of soliton dynamics, whereas the hardening regime is more susceptible to localization loss as uncertainty levels increase. These findings support the feasibility of experimental implementations of solitons in lattices with realistic parametric variability, particularly in applications involving vibration energy harvesting, vibration control, and sensing.

Despite the theoretical findings, several open questions remain. In particular, the influence of the excitation frequency $\Theta$ on both the robustness of the phenomenon and the validity range of the first- and second-order approximations in $\varepsilon$ remains to be established. Although no nucleation has been identified in periodic chains, a more rigorous analytical and numerical assessment is required to draw definitive conclusions. Furthermore, unlike previously reported cases, the governing equation lacks an associated stability diagram, representing an important direction for future research. The effects of nonlinear coupling, as well as the extension of the framework to subharmonic regimes, also warrant further investigation.


\section*{Declarations}

\bmhead*{Acknowledgments}

The authors acknowledge the financial support of the São Paulo Research Foundation (FAPESP) grants 18/15894-0, 2024/22736-3, 25/26998-5; also the Research Foundation - Flanders (FWO) and FAPESP through the bilateral research grant FWO G0F9922N and FAPESP 21/05510-3. The Brazilian authors also acknowledge the support of the Research Agencies CNPq proc. 304932/2024-8 and 406148/2022-8, FINEP under Grant 2091/23, FAPEMIG and CAPES through the INCT-EIE for the financial support provided for this research effort. This work has been supported by the EIPHI Graduate School (Contract No. ANR-17-EURE-0002).

\bmhead*{Conflict of interest}
The authors declare that they have no conflict of interest.

\clearpage
\onecolumn

\begin{appendices}

\newpage
\section{Auxiliary variables employed in the mathematical formulation}\label{Auxiliary variables employed in the mathematical formulation}

\begin{table*}[h]
\centering
\caption{Mathematical relations defining the lattice parameters for periodic mass.}
\label{parâmetros_lattice}
\renewcommand{\arraystretch}{3}

\begin{tabularx}{\linewidth}{l X | l X}
\toprule
\multicolumn{2}{c|}{\makecell{\textbf{Dimensional parameters}\\
\textbf{associated with Equation \eqref{Equação mecânica dim}}}} 
& \multicolumn{2}{c}{\makecell{\textbf{Dimensionless parameters}\\
\textbf{associated with Equation \eqref{Eq mec normalizada}}}  } \\

\midrule
\textbf{Variable} & \textbf{Mathematical relation} &
\textbf{Variable} & \textbf{Mathematical relation} \\
\midrule

$\omega_0^2$ & $= \dfrac{k_l}{m}$ &
$t$ & $= t_f \,\omega_0$ \\

$\zeta_n$ & $ = \dfrac{b_n}{2\omega_0 m}$ &
$u_n(t)$ & $= \dfrac{\sqrt{\beta_{\varepsilon}}\, U_n(t_f)}{\omega_0}$ \\

$f_n$ & $ = \dfrac{\Delta k_{ln}}{m}$ &
$\mu(n)$ & $ = \dfrac{f_{\varepsilon}(n)}{\omega_0^{2}}$ \\

$\beta$ & $ = \dfrac{k_{nl}}{m}$ &
$\Theta$ & $ = \dfrac{\Omega}{\omega_0}$ \\

$C_l^{(k)}$ & $ = \dfrac{k_{cl}^{(k)}}{m}$ &
$\chi_l^{(k)}$ & $= \dfrac{C_{l\varepsilon}^{(k)}}{\omega_0^{2}}, \quad k = 1, \ldots, \alpha$ \\

$C_{nl}^{(k)}$ & $ = \dfrac{k_{cnl}^{(k)}}{m}$ &
$\chi_{nl}^{(k)}$ & $ = \dfrac{C_{nl\varepsilon}^{(k)}}{\beta_{\varepsilon}}, \quad k = 1, \ldots, \alpha$ \\

${F_n}$& $ = \Omega^2U_b$ &
$K_n$ & $= \dfrac{F_n \sqrt{\beta_{\varepsilon}}}{\omega_0^{3}}$ \\

\bottomrule
\end{tabularx}
\end{table*}

\begin{table*}[h]
\centering
\caption{Auxiliary variables employed in Equation \eqref{Seculares_2}.}
\label{auxiliares_sec_2}
\renewcommand{\arraystretch}{4}

\begin{tabularx}{\linewidth}{l X}
\toprule
\textbf{Variable} & \textbf{Mathematical relation} \\
\midrule
 
$\mathcal{L}_1$ &   \makecell[l]{$=
\dfrac{3{\Lambda^{3}_{n}}(\Theta^2-3\Theta+18)}{4\Theta(\Theta-3)}$}   \\
  
 $\mathcal{L}_2$ &   \makecell[l]{$= 
- \dfrac{3{\Lambda^{5}_{n}}(\Theta^2-7)}{2(\Theta^{2}-1)}
\pm\left( \dfrac{ i{\Lambda^{3}_{n}}\zeta_{n\varepsilon}(\Theta^2+12\Theta-1)}{2(\Theta^{2}-1)}
- \dfrac{{\Lambda^{3}_{n}}\sigma}{2}
- \dfrac{{\Lambda^{3}_{n}}\mu(n)(\Theta^2-13)}{4(\Theta^2-1)}
\right)$}   \\

$\mathcal{L}_3$ &   \makecell[l]{$=
- \dfrac{3{\Lambda^{3}_{n}(\Theta^2-19)}}{2(\Theta^2-1)}
$}   \\

$\mathcal{L}_4$ &   \makecell[l]{$=
\pm\dfrac{3{\Lambda^{2}_{n}}\mu(n)(5-\Theta^2)}{\Theta^{2}-1}
- \dfrac{9{\Lambda^{4}_{n}}(2\Theta^2-11)}{2(\Theta^{2}-1)}
- \zeta_{n\varepsilon}^{2}
- \dfrac{\mu^{2}(n)}{4}$}   \\

$\mathcal{L}_5$ &   \makecell[l]{$= 
- \dfrac{9{\Lambda^{2}_{n}}(\Theta^4-22\Theta^2+69)}{(\Theta^{2}-9)(\Theta^2-1)}
\pm\left(- 3i\zeta_{n\varepsilon}- \dfrac{3\mu(n)}{2}\right)$}   \\\\

$\mathcal{F}_{\chi_l^{(k)}}$ &
\makecell[l]{
$=
\sum_{k=1}^{\alpha}\chi_l^{(k)}
\Bigg[
\pm\Bigg(
\dfrac{3A_n^2\Delta_k\{\overline{A}\}}{4}
-\dfrac{3\Delta_k\{A\left|A\right|^2\}}{4}
-\dfrac{3\left|A_n\right|^2\Delta_k\{A\}}{2}
$ \\
$\qquad
-3\Lambda_n^2\Delta_k\{A\}
\Bigg)
-\dfrac{\mu(n)\Delta_k\{A\}}{4}
-\dfrac{\Delta_k\{\mu A\}}{4}
\Bigg]
$
} \\

$\mathcal{F}_{\chi_l^{(k)}\chi_l^{(j)}}$ &   \makecell[l]{$= \sum_{k=1}^{\alpha}\sum_{j=1}^{\alpha} \chi_l^{(k)}\chi_l^{(j)} \left[
\dfrac{\Delta_{|k-j|}\{A\}}{4}
+ \dfrac{\Delta_{k+j}\{A\}}{4} 
- \dfrac{\Delta_{j}\{A\}}{2}
- \dfrac{\Delta_{k}\{A\}}{2}
\right]$} \\

$\mathcal{F}_{\chi_{nl}^{(k)}}$ &   \makecell[l]{$= \sum_{k=1}^{\alpha} \chi_{nl}^{(k)} \left[ 3\left(A_n-A_{n+k}\right)^2\left(\overline{A_n} - \overline{A_{n+k}}\right) + 3\left(A_n-A_{n-k}\right)^2\left(\overline{A_n} - \overline{A_{n-k}}\right)\right]$} \\

${\Delta_j}\{X\}$ &   \makecell[l]{$= 2X_{n} - X_{n+j} - X_{n-j}$ \, \, \, \, for any expression $X$} \\

\bottomrule
\end{tabularx}
\end{table*}

\begin{table*}[h]
\centering
\caption{Auxiliary variables employed in Equations \eqref{seno_fase} and \eqref{coseno_fase} .}
\label{auxiliares_sen_coseno}
\renewcommand{\arraystretch}{4}

\begin{tabularx}{\linewidth}{l X}
\toprule
\textbf{Variable} & \textbf{Mathematical relation} \\
\midrule
 
$V$ &  \makecell[l]{$= \dfrac{a_{n}\,\varepsilon\,{\zeta_n}_{\varepsilon}\left(\pm3a_{n}^{2}\varepsilon - 8\right)}{8}$} \\

$W$ &  \makecell[l]{$= \dfrac{3\Lambda_n^{5}\varepsilon^{2}\left(\Theta^{2}-7\right)}{2(\Theta^{2}-1)}
+
\dfrac{3\Lambda_n^{3}\varepsilon^2a_{n}^{2}\left(\Theta^{2} - 19\right)}{8(\Theta^{2}-1)}
\pm \dfrac{\Lambda_n^{3}(\varepsilon^2\sigma-2\varepsilon)}{2} $}  \\

$X$ &  \makecell[l]{$= \pm\dfrac{\Lambda_n^{3}\varepsilon^{2}{\zeta_n}_{\varepsilon}\left(-\Theta^{2}-12\Theta+1\right)}{2(\Theta^{2}-1)}$}  \\

$Y$ &  \makecell[l]{$= -\dfrac{3  a_{n}^{2} \Lambda_n^{3} \varepsilon^{2} \left(\Theta^{2} - 3\Theta + 18\right)}{16\,\Theta\left(\Theta - 3\right)}$}  \\

$Z$ & \makecell[l]{
$\displaystyle
= \dfrac{15 a_{n}^{5} \varepsilon^{2}}{256}
\pm\dfrac{3\varepsilon a_n(-a_n^2-8\Lambda_n^2)}{8}
+ \dfrac{9\varepsilon^2 a_n^3\Lambda_n^2(\Theta^4-22\Theta^2+69)}
{8(\Theta^2-9)(\Theta^2-1)}
$ \\
$\displaystyle
+ \dfrac{9\varepsilon^2 a_n\Lambda_n^4(2\Theta^2-11)}
{4(\Theta^2-1)}
+\dfrac{\varepsilon^2 \zeta_\varepsilon^2 a_n}{2}
$
} \\

\bottomrule
\end{tabularx}
\end{table*}

\begin{table*}[h]
\centering
\caption{Auxiliary variables employed in Equation \eqref{E_c}.}
\label{tabela Ec}
\renewcommand{\arraystretch}{4}

\begin{tabularx}{\linewidth}{l X}
\toprule
\textbf{Variable} & \textbf{Mathematical relation} \\
\midrule
 
$\Pi_{1}$ &  
\makecell[l]{$\displaystyle 
= \pm\dfrac{3 a_{n} \Lambda_n^{2} \sin(\gamma_{n})}{4 \Theta (\Theta - 1)} 
- \dfrac{4 \Theta \Lambda_n \zeta_{\varepsilon}}{\Theta^{2} - 1}$}  \\

$\Pi_{2}$ &  
\makecell[l]{$\displaystyle 
= \pm\dfrac{3 a_{n} \Lambda_n^{2} \cos(\gamma_{n})}{4 \Theta (\Theta - 1)} 
+ \dfrac{\pm3 \Lambda_n \left(a_{n}^{2} + 2 \Lambda_n^{2}\right) + 2 \Lambda_n \mu(n)}{\Theta^{2} - 1}$}  \\

\bottomrule
\end{tabularx}
\end{table*}

\clearpage
\newpage
\section{Supplementary numerical simulations}\label{Supplementary numerical simulations}

\begin{figure*}[ht!]
    \centering
    \includegraphics[width=1\textwidth]{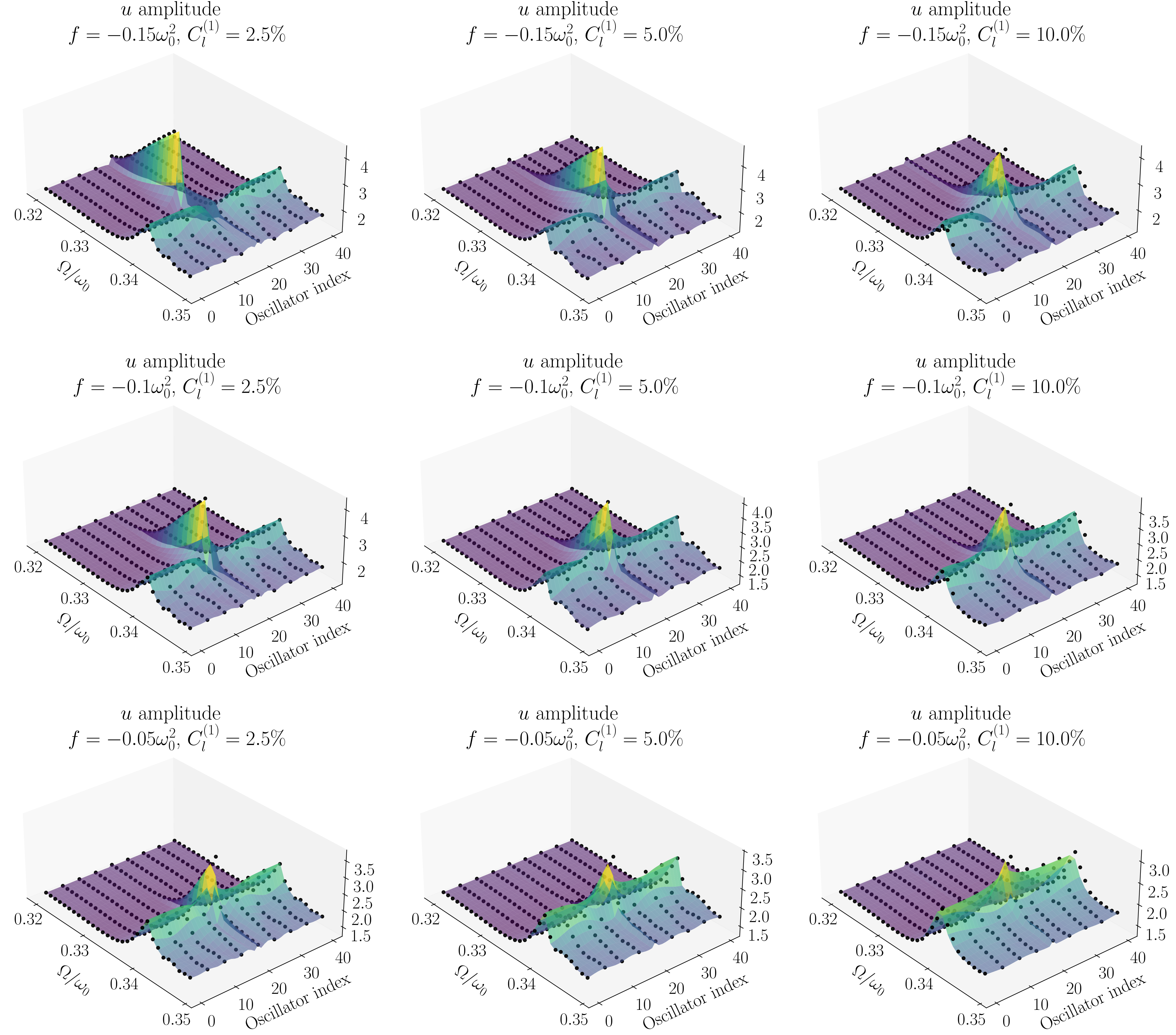}
    \caption{Numerical validation of the analytical solution given by Eq.~\eqref{soliton}. Black markers correspond to the numerical integration of the lattice equation~\eqref{Eq mec normalizada}, while the colored surface represents the analytical solution. The point density was reduced for visualization purposes. Hardening regime ($\pm = 1$). The analytical solution is obtained using the DOP853 method, whereas the numerical integration of the lattice equation is performed using the LSODA method, both with tolerances $rtol = 10^{-6}$ and $atol = 10^{-8}$. The adopted parameters are $\zeta = 0.3/100$, $\omega_0 = 100\,\mathrm{Hz}$, $\beta = 2.5 \times 10^9\,\mathrm{N/(kg \cdot m^3)}$, $e = 0.01$, and $F = 700\,\mathrm{m/s^2}$. Nonlinear coupling is neglected, while linear coupling is considered up to third order ($\alpha = 3$), with coefficients $C_l^{(2)} = C_l^{(1)}/4$ and $C_l^{(3)} = C_l^{(1)}/9$. All parameters are periodic except for the introduced disorder. Numerical continuation is employed for the analytical solution.}
    \label{validação_soliton_num}
\end{figure*}

 \begin{figure*}[h]
    \centering
    \includegraphics[width=1\textwidth]{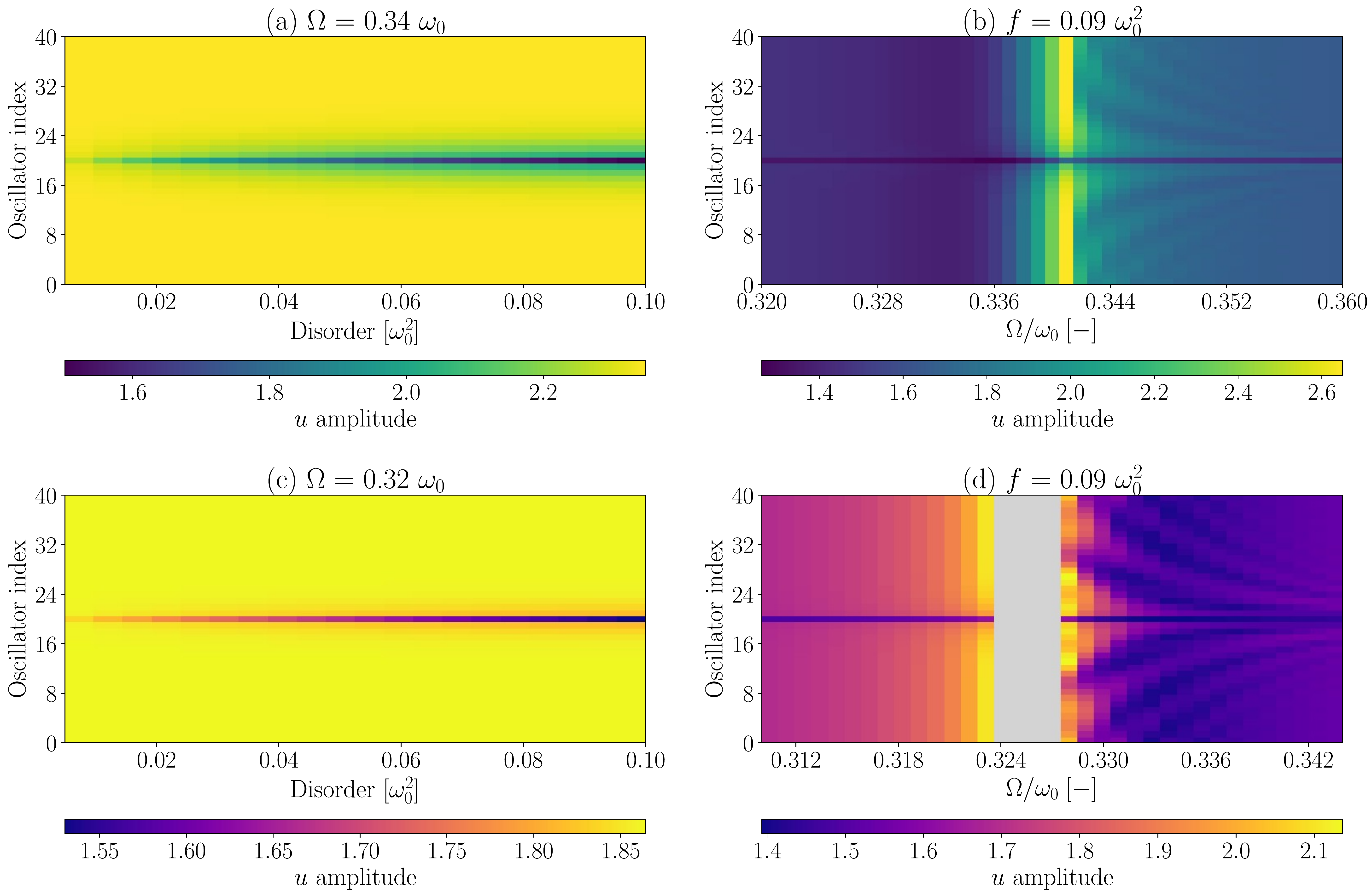}
    \caption{Numerical sweeps of soliton nucleation for hardening (a–b) and softening (c–d) regimes under positive values of $f(n)$. Panels (a,c): fixed excitation frequency with varying disorder; (b,d): fixed disorder with varying frequency. Simulations use the DOP853 method ($rtol=10^{-6}$, $atol=10^{-8}$) over 100 periods. Parameters: $\zeta=0.3/100$, $\omega_0=100$ Hz, $\beta=2.5\times10^9\mathrm{N/(kg\cdot m^3)}$, $e=0.01$, and $F=700\mathrm{m/s^2}$. Gray region (c–d): no steady-state convergence. Absence of nonlinear coupling. Linear coupling up to third order ($\alpha=3$), with $C_l^{(1)} = 2.5\%,\omega_0^2$, $C_l^{(2)} = C_l^{(1)}/4$, and $C_l^{(3)} = C_l^{(1)}/9$. Periodic parameters, except for the introduced disorder. Numerical continuation imposed on the frequency domain.}
    \label{prova_de_solitonIMPpos}
\end{figure*}

 \begin{figure*}[h]
    \centering
    \includegraphics[width=1\textwidth]{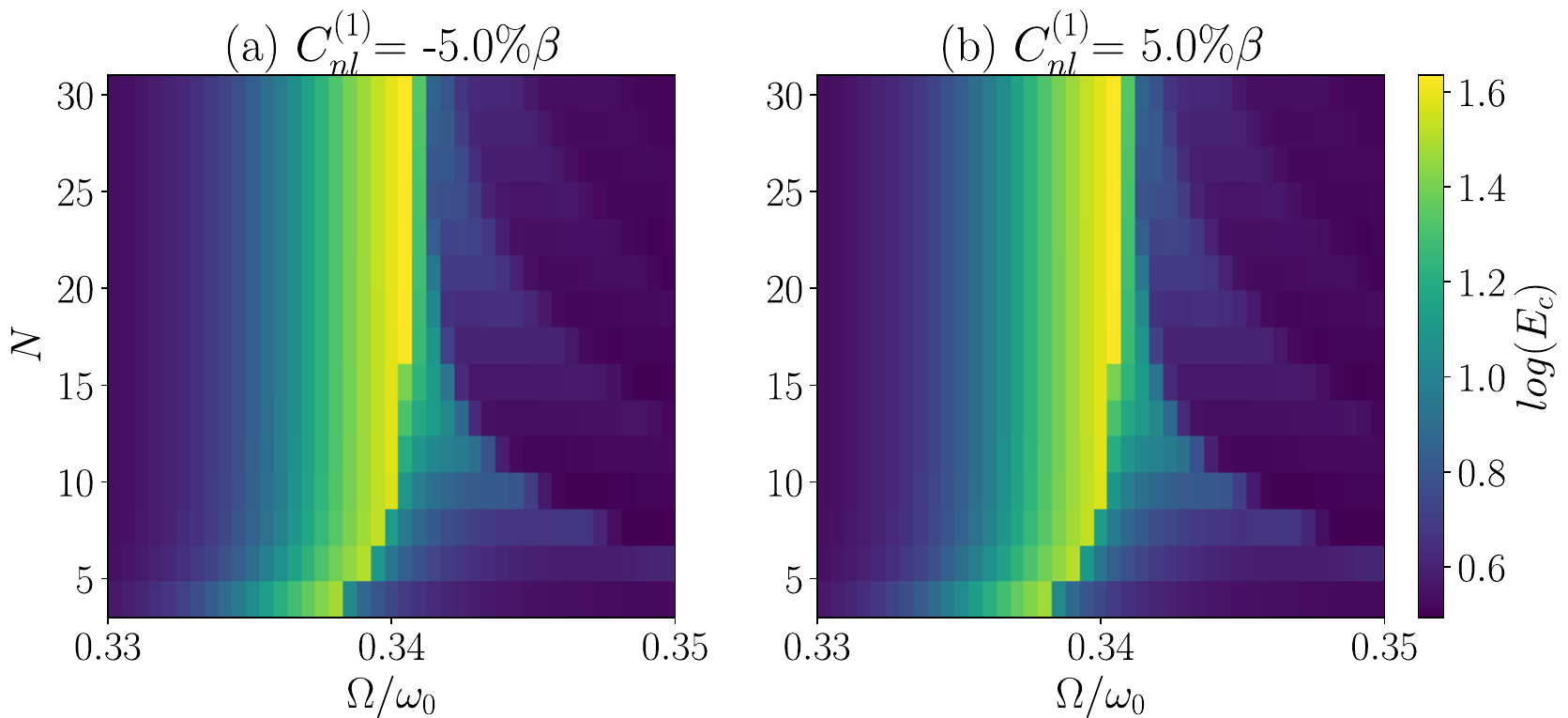}
    \caption{Numerical related to the nonlinear coupling ($\pm = 1$). Simulations are performed using the DOP853 method with tolerances $rtol = 10^{-6}$ and $atol = 10^{-8}$. The adopted parameters are $\zeta = 0.3/100$, $\omega_0 = 100\,\mathrm{Hz}$, $\beta = 2.5 \times 10^9\,\mathrm{N/(kg \cdot m^3)}$, $e = 0.01$, and $F = 700\,\mathrm{m/s^2}$. Linear coupling is considered up to third order ($\alpha = 3$), with coefficients $C_l^{(2)} = C_l^{(1)}/4$ and $C_l^{(3)} = C_l^{(1)}/9$ and $C_l^{(1)} = 2.5\%,\omega_0^2$. All parameters are periodic except for the introduced disorder ($f = -0.05\omega_0^2$). Numerical continuation is not employed.}
    \label{ParamKnlNH}
\end{figure*}

 \begin{figure*}[h]
    \centering
    \includegraphics[width=1\textwidth]{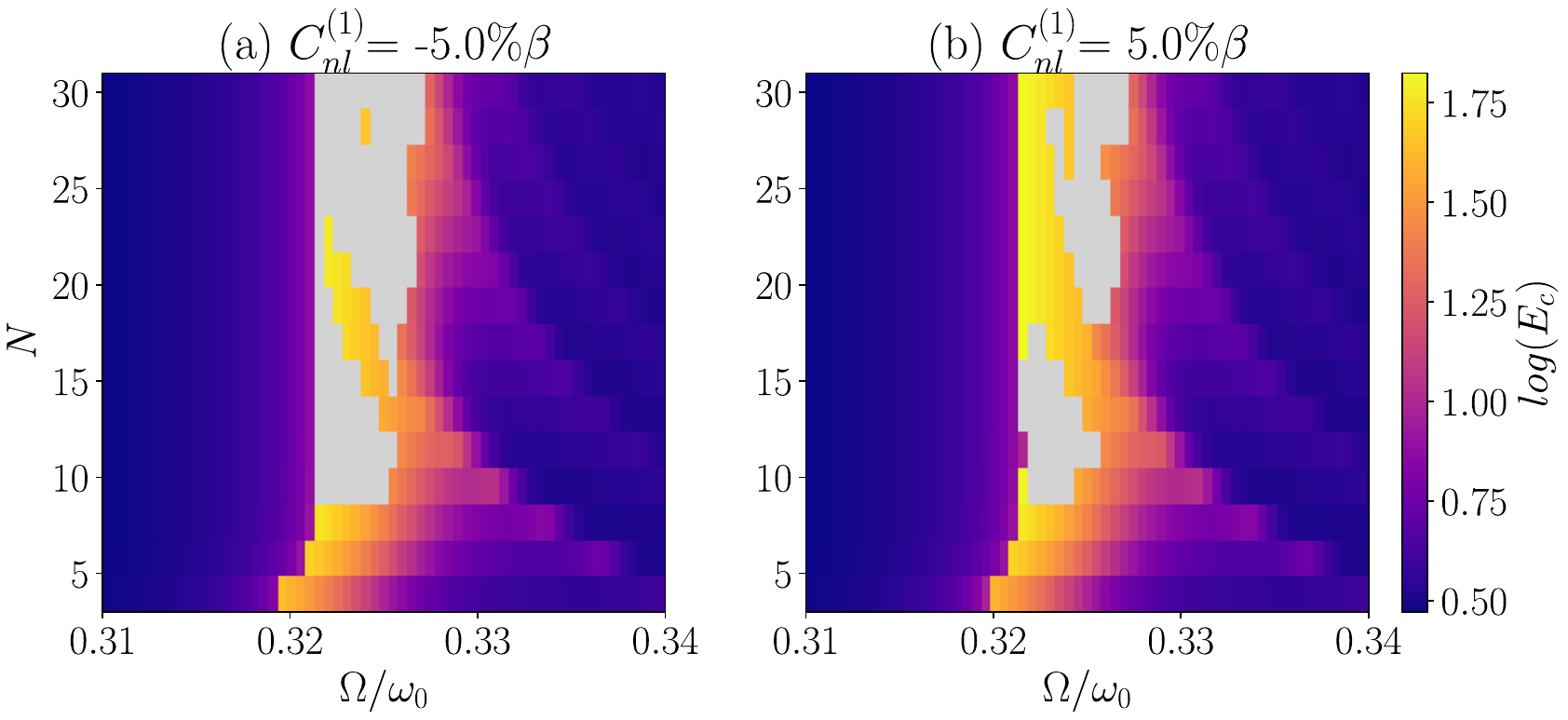}
    \caption{Numerical related to the nonlinear coupling ($\pm = -1$). Simulations are performed using the DOP853 method with tolerances $rtol = 10^{-6}$ and $atol = 10^{-8}$. The adopted parameters are $\zeta = 0.3/100$, $\omega_0 = 100\,\mathrm{Hz}$, $\beta = 2.5 \times 10^9\,\mathrm{N/(kg \cdot m^3)}$, $e = 0.01$, and $F = 700\,\mathrm{m/s^2}$. Linear coupling is considered up to third order ($\alpha = 3$), with coefficients $C_l^{(2)} = C_l^{(1)}/4$ and $C_l^{(3)} = C_l^{(1)}/9$ and $C_l^{(1)} = 2.5\%,\omega_0^2$. All parameters are periodic except for the introduced disorder ($f = -0.05\omega_0^2$). Numerical continuation is not employed.}
    \label{ParamKnlNS}
\end{figure*}

 \begin{figure*}[h]
    \centering
    \includegraphics[width=1\textwidth]{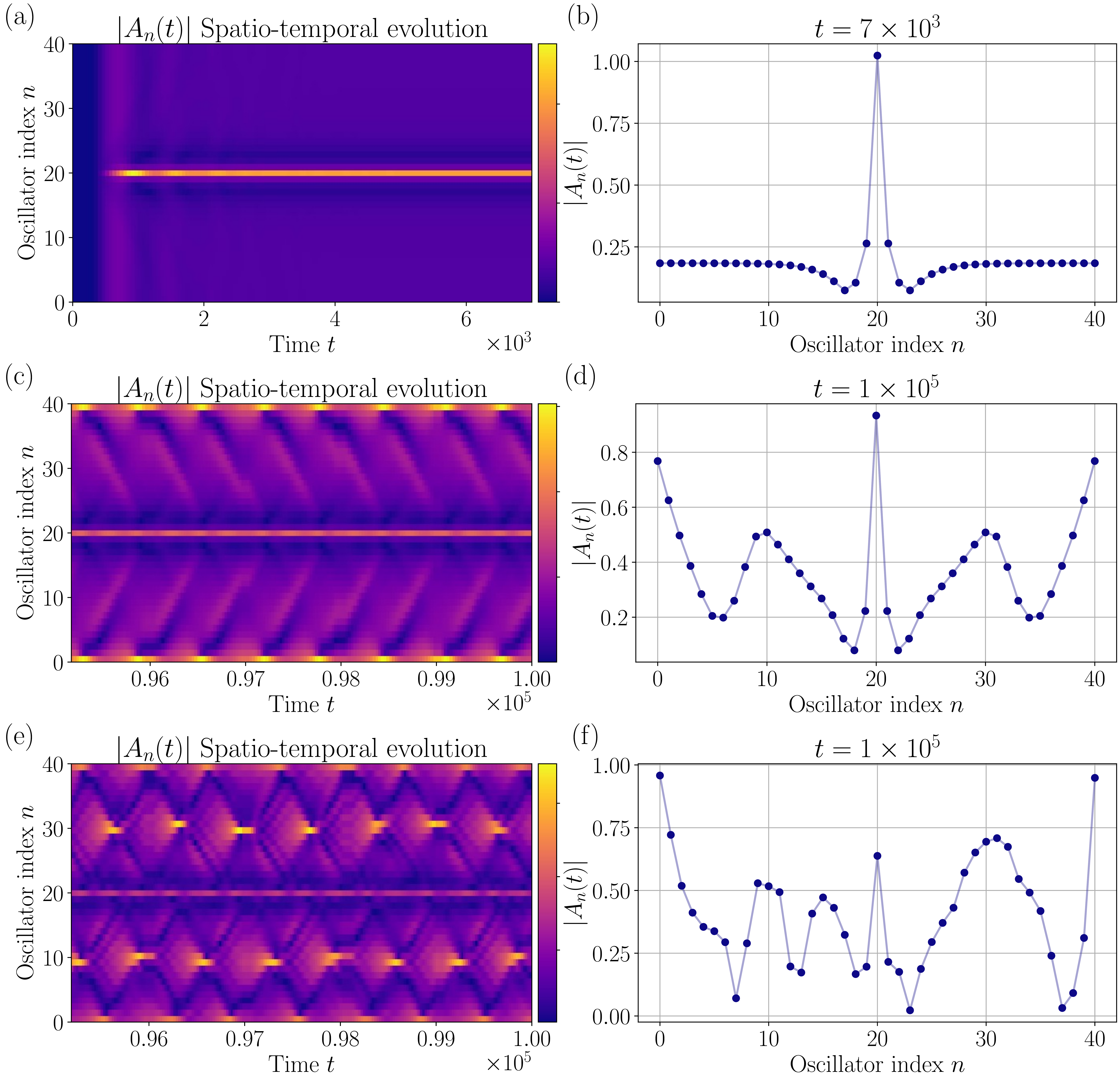}
    \caption{Space--time evolution obtained from Eq.~\eqref{soliton}. Panel (a) presents a temporal response satisfying the convergence criterion defined in Eq.~\eqref{critério de regime permanente}, whereas panels (c) and (e) do not satisfy this condition. Panel (c) exhibits a non-convergent periodic response, while panel (e) displays a non-periodic behavior. Softening regime ($\pm = -1$). Panels (b), (d), and (f) detail the corresponding amplitude profiles after the transient regime. The convergent response shown in (a)--(b) is obtained for $\Theta = 0.322$, the non-convergent periodic response in (c)--(d) for $\Theta = 0.324$, and the non-convergent chaotic response in (e)--(f) for $\Theta = 0.326$. Time integration is performed using the DOP853 method with tolerances $rtol = 10^{-6}$ and $atol = 10^{-8}$. The adopted parameters are $\zeta = 0.3/100$, $\omega_0 = 100\,\mathrm{Hz}$, $\beta = 2.5 \times 10^9\,\mathrm{N/(kg \cdot m^3)}$, $e = 0.01$, $F = 700\,\mathrm{m/s^2}$, and $C_l^{(1)} = 2.5\%\,\omega_0^2$. Linear coupling is considered up to third order ($\alpha = 3$), with coefficients $C_l^{(2)} = C_l^{(1)}/4$ and $C_l^{(3)} = C_l^{(1)}/9$. Nonlinear coupling is neglected. All parameters are periodic except for the introduced disorder ($f = -0.05\,\omega_0^2$).}
    \label{Timeevo}
\end{figure*}

\end{appendices}

\clearpage

\end{document}